\documentclass{article}
\usepackage[utf8]{inputenc}
\usepackage[a4paper, total={6.25in, 9.5in}]{geometry}
\usepackage{rotating}

\usepackage{microtype}
\usepackage{booktabs} 

\usepackage{amsmath, amssymb, amsfonts, amsthm, bbm}
\usepackage{algorithm, algpseudocode, algorithmicx}
\usepackage{mathtools,commath}
\usepackage{graphicx, multirow}
\usepackage[round]{natbib}
\usepackage{xcolor}
\usepackage{authblk}
\usepackage{hyperref}	
\usepackage[capitalise]{cleveref}
\usepackage{comment}

\usepackage{caption}
\usepackage{subcaption}
\usepackage[section]{placeins}

\theoremstyle{plain}

\theoremstyle{plain}

\usepackage{algorithm,algpseudocode}

\makeatother

\usepackage{babel}
\providecommand{\algorithmname}{Algorithm}
\providecommand{\theoremname}{Theorem}

\usepackage[title]{appendix}

\usepackage{setspace}
\begin{document}
\title{Computationally Efficient Estimation of Large Probit Models
}
\date{\today}
\author{Patrick Ding\thanks{\scriptsize Microsoft Corporation, \url{patrickding00@gmail.com}}  \quad
Guido Imbens\thanks{\scriptsize Stanford University, Department of Economics and Graduate School of Business, \url{imbens@stanford.edu}} \quad
Zhaonan Qu\thanks{\scriptsize Corresponding author. Columbia University, Data Science Institute, \url{zq2236@columbia.edu}} \quad Yinyu Ye\thanks{\scriptsize Stanford University, Department of Management Science and Engineering, \url{yyye@stanford.edu}}}
\maketitle
\begin{abstract}
Probit models are useful for modeling correlated discrete responses in many disciplines, including consumer choice data in economics and marketing. 
However, the Gaussian latent variable feature of probit models coupled with identification constraints pose significant computational challenges for its estimation and inference, especially when the dimension of the discrete response variable is large.
In this paper, we propose a computationally efficient Expectation-Maximization (EM) algorithm for estimating large probit models. Our work is distinct from existing methods in two important aspects. First, instead of simulation or sampling methods, we apply and customize expectation propagation (EP), a deterministic method originally proposed for approximate Bayesian inference, to estimate moments of the
truncated multivariate normal (TMVN) in the E (expectation) step. Second, we take advantage of a symmetric identification condition to transform the constrained optimization problem in the M (maximization) step into a one-dimensional problem, which is solved efficiently using
Newton's method instead of off-the-shelf solvers. Our method enables the analysis of correlated choice data in the presence of more than 100 alternatives, which is a reasonable size in modern applications, such as online shopping and booking platforms, but has been difficult in practice with probit models. We apply our probit estimation method to study ordering effects in hotel search results on Expedia's online booking platform.
\end{abstract}
{\it Keywords:}  Expectation Maximization; Expectation Propagation; Newton's Method; Discrete Choice 
\section{Introduction}
\label{sec:intro}
Correlated discrete responses are common across social sciences, statistics, computer science, and
life sciences. For example, in consumer choice data, we observe
each customer's choice over a set of alternatives, which could be correlated with each other through substitution effects. Another popular setting is panel and time series data \citep{greene2004convenient,bertschek1998convenient}. For example, in epidemiology,
one may have time series observations of binary indicators of symptoms for patients, and 
periods close to each other may exhibit stronger serial correlation \citep{zhang2000model}. Stock price changes are also often measured in discrete ``ticks'' over time periods \citep{hausman1992ordered}. It is therefore important to accurately model and estimate the correlations in data with multi-dimensional discrete response variables.

Probit \citep{thurstone1927method} and logit \citep{luce2012individual,berkson1944application} are two classes of latent variable models widely used for data with discrete or categorical responses. 
However, their distinct features have led to vastly
different developments and adoptions in the past decades. In particular, because of the computational convenience  afforded by
its tractable likelihood, logit type models have found widespread applications ranging
from designing transportation systems to powering workhorse tool-kits
in modern machine learning. Despite this success, in applications with correlated response variables, logit type models have important limitations. For example, in discrete choice modeling, the multinomial logit model cannot adequately capture correlations due to the independence of irrelevant
alternatives (IIA) property. Although extensions such as the nested logit model \citep{ben1973structure} and mixed logit model \citep{boyd1980effect,mcfadden2000mixed, berry1995automobile} have been proposed to relax this restriction in discrete choice settings, the model specification often requires domain knowledge
about the problem at hand, and may also preclude discovery of possible
correlations unknown to the researcher. 

In this regard, probit models possess one
major advantage over logit models. The covariance matrix of latent Gaussian utilities in a probit model can accommodate general correlation
structures within the discrete response variable, which can account for contextual effects not captured by logit models and capture general substitution patterns beyond the IIA restrictions in discrete choice \citep{train2009discrete}.  Despite this desirable feature, probit models suffer from considerable computational challenges due to the complicated form of its likelihood function, which integrates truncated multivariate normal (TMVN) densities and does not have a closed-form expression. This challenge is further compounded by the identification conditions of probit models. These difficulties are a main reason that logit type models, such as the mixed logit, are much more popular in practice for modeling choice behavior. 

Previous works have proposed several approaches to improve the computational efficiency of probit estimation. Many rely on Bayesian analysis \citep{chib1998analysis,mcculloch2000bayesian,burgette2012trace,loaiza2021scalable} and/or simulations \citep{mcfadden1989method,hajivassiliou1998method,natarajan2000monte}. Among the most popular simulation-based approaches for multinomial probit is the GHK sampler \citep{geweke1989bayesian,hajivassiliou1996simulation,keane1994computationally}. Although they have proven to be  efficient and accurate in practice, the application of these approaches for probit estimation have largely been limited to data with small dimensions in the discrete response variable, e.g., choice sets with size less than 10. On the other hand, modern applications involving discrete choice, such as online shopping and booking platforms, are rapidly developing and expanding in size, with potentially as many as a few hundred alternatives as a result of product differentiation. Importantly, in such applications we are often interested in the correlation structure or contextual effects in the data, which makes probit models a suitable choice for data analysis. However, existing methods are either too slow or unable to handle the estimation of large probit models on such datasets. It is therefore of practical value to have a computationally efficient probit estimation method that can reliably handle large response variable dimensions/choice sets.

In this paper, we propose the EP-Newton EM algorithm, a computationally efficient probit estimation method that can handle data with large dimensional response variables. Our \emph{deterministic} 
 algorithm employs the Expectation Propagation (EP) procedure \citep{minka2001expectation,minka2001family} 
 and does not require simulating choice probabilities or drawing samples from TMVN. Our work is distinct from existing works in three main aspects. First, although Bayesian methods have generally been preferred for probit estimation due to their superior empirical performance on moderate sized choice sets \citep{chib1998analysis,mcculloch2000bayesian,loaiza2021scalable,loaiza2023fast}, they require non-standard priors on the covariance/precision matrix due to probit model's identification (normalization) constraints, which leads to additional computational complications for large choice sets. We circumvent this issue by taking an EM approach, which handles identification in the constrained optimization problem in the M (maximization) step. Second, in sharp contrast to existing EM approaches for probit estimation, we propose to use the deterministic expectation propagation (EP) procedure to directly compute the conditional moments of the TMVN in the E (expectation) step \citep{cunningham_gaussian_2013}. 
 This approach avoids computationally expensive sampling from TMVNs.
 In addition, our approach allows for the possibility to customize and accelerate the EP method specifically for probit models by leveraging the sparsity and involution structures of the constraint matrix. 
 Third, current EM approaches usually use off-the-shelf solvers for the constrained optimization problem in the M step, which may be inefficient from a computational perspective, especially when the problem is subject to \emph{asymmetric} identification constraints commonly adopted by previous works on probit estimation. Instead, we propose to use a symmetric trace constraint on the precision (inverse covariance) matrix to enforce the scale identification condition. This novel identification restriction allows us to transform the constrained optimization problem into a \emph{one-dimensional} fixed point problem, which can be solved efficiently with a simple application of Newton’s method. Our symmetric identification condition is inspired by the work of \citet{burgette2012trace}, who propose a symmetric trace constraint on the \emph{covariance} matrix in the context of Bayesian probit estimation. Our proposal therefore also demonstrates that the choice of identification constraint is important for the optimization of likelihood downstream. Our proposed probit estimation framework can be readily extended beyond the mutli-dimensional probit models discussed in this paper to other probit specifications, such as \citet{hausman1992ordered}, and in addition can handle data with \emph{varying} choice set sizes, which is almost always the case in real world applications.

 Numerical studies demonstrate that our innovations combine to significantly improve the computational efficiency of probit estimation. On small to medium sized problems, our estimation method achieves comparable or better accuracy compared to existing methods, but requires orders of magnitude less computation time. On large problems (dimension larger than 100), our method can obtain good parameter estimates within reasonable time, while previous methods fail or cannot converge due to their intensive computational burdens on data with large dimensional response variables. We further improve the computational efficiency with sub-sampling and parallelization. An \texttt{R} package is currently under construction and available on Github\footnote{Package available on Patrick Ding's Github at \url{https://github.com/delimited0/fsprobit}.}, and we believe it could help researchers efficiently estimate large probit models on data with high-dimensional correlated discrete responses. 

To further demonstrate the usefulness of our approach, we apply our probit estimation approach to the hotel search and booking data on Expedia.com\footnote{Available at \url{https://www.kaggle.com/competitions/expedia-personalized-sort/data}.} where results are displayed in a ranked list, in order to study the \emph{heterogeneous} effects of higher ranks on increases in demand for different hotels and ordered choice sets. Such effects cannot be easily captured using logit type models, which would require a large number of dummy variables of position. We find that the conditional correlations between the random utilities of the top positions in a ranked list are positive, while others are negative. 
 Moreover, the top position is unique in that the marginal correlations of its random utility with the 2nd to 5th positions are relatively smaller than the other marginal correlations. 
 In addition, we find heterogeneous (in property characteristics) increases in demand based on predictions made by our estimated probit model.
 The predicted increases in demand are positively correlated with the original position of properties, but only up to the first fifteen or so positions.
 Afterwards, the benefit of sponsored spots stays mostly constant. 
 These observations are well aligned with a limited attention framework of decision-making \citep{caplin2019rational,abaluck2021consumers}, and suggest that sponsored spots are indeed an important tool for merchants to boost the demands for their properties and products. 
 Moreover, because of the heterogeneous benefits of sponsored spots for different properties, the platform could potentially capture more profit and provide better service by employing a more \emph{individualized} pricing structure for such sponsored spots. 

The rest of the paper is organized as follows. 
We discuss the probit maximum likelihood estimation problem and the general EM framework for it in \cref{sec:setup}. We then provide a survey of existing works in \cref{sec:literature-review}. We propose our EP-Newton EM algorithm for probit estimation in \cref{sec:EP-Newton-EM}.
In \cref{sec:simulations} we conduct simulation studies, and in \cref{sec:expedia} we apply our method to the Expedia dataset. 
\section{Maximum Likelihood Estimation of Probit Models}
\label{sec:setup}
In this section, we first provide a review of the most common multi-dimensional probit models, namely the multivariate and multinomial probit models, and set up their maximum likelihood estimation problems. We then discuss the general expectation maximization (EM) framework for their maximum likelihood estimation. Our main contributions, detailed in \cref{sec:E-step,sec:M-step}, provide a computationally efficient algorithm based on the EM framework that is distinct from existing algorithms for probit models. 
\subsection{The Multivariate and Multinomial Probit Models}
Both probit \citep{thurstone1927method,bliss1934method,fisher1935case} and logit models \citep{luce2012individual,berkson1944application} were initially developed for one-dimensional
binary responses. For completeness, in \cref{sec:probit-logit-introduction} we provide an introduction of the one-dimensional models and their latent variable interpretations. The models are readily extended to multi-dimensional discrete responses. More precisely, in this paper we assume that in each observation we have a binary response vector $Y\in\{0,1\}^{m}$ which can represent different types of discrete response variables. In discrete choice data over $m$ alternatives, the response variable is the choice among $m$ alternatives, which is equivalently represented as an element in $\{0,1\}^{m}$, with the entry corresponding to the chosen alternative being 1 and all others being 0. 
In panel data with $m$ periods, each response variable corresponds to a sample's time-indexed outcomes. For example, in epidemiology $Y$ could be the indicators of whether a patient experiences a particular symptom in each of the $m$ time periods. In this case, each entry of $Y$ is a binary indicator corresponding to one of $m$ time periods,
and $Y$ can have multiple entries equal to 1. 

Suppose now for each observation $i\in\{1,\dots,n\}$ we observe the response variable $Y_{i}\in\{0,1\}^{m}$ and the matrix of covariates $X_{i}\in\mathbb{R}^{m\times p}$, where the $j$-th row of $X_{i}$, denoted by $X_{ij}\in\mathbb{R}^{p}$, corresponds to the $j$-th dimension of the $i$-th observation-specific covariates.
In our multi-dimensional probit models, we specify a latent variable $Z_{i}\in\mathbb{R}^{m}$ for each observation with Gaussian distribution
\begin{align*}
    Z_{i} & \sim\mathcal{N}(X_{i}\beta,\Sigma),
\end{align*}
with unknown coefficient vector $\beta\in\mathbb{R}^{p}$ and covariance matrix $\Sigma$. 

In panel data, one common specification of the relation between the observed response variable $Y_{i}$ and the unobserved latent variable $Z_{i}$ is 
\begin{align}
\label{eq:multivariate-probit}    Y_{ij} & =
    \begin{cases}
        1 & \text{if }Z_{ij}\geq0\\
        0 & \text{otherwise}
    \end{cases}.
\end{align}
 If we let $f(Z\mid X,\beta,\Sigma)$ be the PDF of the multivariate Gaussian $\mathcal{N}(X\beta,\Sigma)$, then the likelihood of this probit model is given by 
\begin{align*}
    \prod_{i}\mathbb{P}(Y_{i}\mid X_{i},\beta,\Sigma) & =\prod_{i}\int_{\mathcal{A}_{i}}
    f(Z_{i}\mid X_{i},\beta,\Sigma)dZ_{i1}\dots dZ_{im},
\end{align*}
where $i=1,\dots,n$ indexes the independent observations, $j=1,\dots,m$ indexes the $m$ categories. $\mathcal{A}_i\subseteq \mathbb{R}^m$ is the rectangular region  $\mathcal{A}_i:=A_{i1}\times \cdots \times A_{im}$, where each 
$A_{ij}$ is the interval $(0,\infty)$ if $Y_{ij}=1$ and $(-\infty,0]$ otherwise.
This specification is known as the \textbf{multivariate probit model}. 

On the other hand, in the context of discrete choice modeling, $Z_{ij}$'s are interpreted
as random utilities of alternatives $j$ in observation $i$, and the
observed choices are results of the utility maximizer's rational decision.
Thus in a probit model of discrete choice, $Y_{ij}=1$ if and only
if the $j$-th alternative is selected in the $i$-th observation,
in which case the latent components of the model satisfy 
\begin{align*}
Z_{ij} \geq Z_{ij'}\Longleftrightarrow (X_{i}\beta)_j+\epsilon_{ij} \geq (X_{i}\beta)_{j'}+\epsilon_{ij'},\forall j',
\end{align*}
 leading to the following specification known as the \textbf{multinomial probit model}:
\begin{align}
\label{eq:multinomial-probit}
Y_{ij} & =\begin{cases}
1 & \text{if }Z_{ij}=\max_{j}Z_{ij}\\
0 & \text{otherwise}
\end{cases}.
\end{align}
The likelihood of the multinomial probit model is given by 
\begin{align*}
\prod_{i}\mathbb{P}(Y_{i}\mid X_{i},\beta,\Sigma) & =\prod_{i}\int_{\mathcal{A}_{i}}f(Z_{i}\mid X_{i},\beta,\Sigma)dZ_{i1}\dots dZ_{im},
\end{align*}
where $\mathcal{A}_{i}\subseteq \mathbb{R}^m$ is now the \emph{polytope} region of $z \in \mathbb{R}^m$ with $\arg \max_j z_j = \arg \max_j Y_{ij}$. 

The multivariate probit is often used to model time series and
panel data with discrete response variables, while the multinomial probit
is almost always associated with discrete choice models. For this
reason, we sometimes refer to the former as ``panel probit models'',
and the latter as ``discrete choice probit models''. Other
specifications and extensions of the probit model have also been studied. For example, \citet{hausman1992ordered}
 propose an ``ordered probit'' model for discrete stock price
changes. Their probit specification is given by 
\begin{align*}
Y_{ij} & =\begin{cases}
s_{1} & \text{if }\alpha_{0}\leq Z_{ij}<\alpha_{1}\\
\vdots & \vdots\\
s_{\ell} & \text{if }\alpha_{\ell-1}\leq Z_{ij}<\alpha_{\ell}
\end{cases},
\end{align*}
which generalizes the panel probit model to the case when the
response variable takes values in $\{0,1,\dots,\ell\}^{m}$ rather than
$\{0,1\}^{m}$. Notably, the method we propose in this paper applies generally to all variants of the probit model. The key difference 
lies in how conditional moments of the
truncated multivariate normals are approximated by the EP procedure.

Note that the probit models above parameterized by $(\beta,\Sigma)$ are \emph{invariant} under multiplication by a positive constant $c>0$, in the sense that likelihood and model predictions remain unchanged for $(c\beta,c^2 \Sigma)$. We next discuss in more detail the identification requirements for probit models, which have important statistical and computational implications. 

\subsection{Identification and Estimation of Multi-dimensional Probit Models}
\label{subsec:identification}
Given $n$ samples of data $\{(X_{i},Y_{i})\}_{i=1}^{n}$, 
our task is to obtain the maximum likelihood estimates of $\beta$
and $\Sigma$ in the multivariate probit \eqref{eq:multivariate-probit} or multinomial probit \eqref{eq:multinomial-probit} by solving the optimization problem 
\begin{align}
\label{eq:mle}
\max_{\beta,\Sigma \succ 0, s(\Sigma)=0} -\frac{n}{2}\log\det(\Sigma)+\sum_{i}\log\int_{\mathcal{A}_{i}}\exp(-\frac{1}{2}(Z_{i}-X_{i}\beta)^{T}\Sigma^{-1}(Z_{i}-X_{i}\beta))dZ_{i},
\end{align}
where $s(\Sigma)=0$ is some normalization constraint described shortly, and $\mathcal{A}_{i}$ is the appropriate region of integration for specific probit models.

Note first that the log-likelihood function in \eqref{eq:mle} is not concave in either $\Sigma$ or $\Sigma^{-1}$: the first term is
convex in $\Sigma$ and concave in $\Sigma^{-1}$, while the second
is only convex in $\Sigma^{-1}$ because of the log-sum-exp function. This feature is in contrast
to the standard least squares regression with unknown covariance, where the second term is simply $-\sum_{i}\frac{1}{2}(Z_{i}-X_{i}\beta)^{T}\Sigma^{-1}(Z_{i}-X_{i}\beta)$
and is hence linear in $\Sigma^{-1}$, resulting in a log-likelihood function that is jointly concave in $\beta$ and in $\Sigma^{-1}$. Thus the maximum likelihood estimation of probit models is a non-convex optimization problem, and could have multiple
local optima. More importantly, probit models require additional conditions in order to guarantee identification. Before we discuss how to solve \eqref{eq:mle} in detail, we first address the two main identification issues of probit models. 

\textbf{Scale Invariance.} First of all, for both the multivariate and multinomial probit models, the distribution on data is invariant under multiplication of $Z$ by a positive constant $c$. This ``scale invariance'' of probit models can be addressed by a normalization constraint, represented as $s(\Sigma)=0$ in \eqref{eq:mle} for some $s$. For example, a popular choice in previous works is to require $\Sigma_{11}=1$ \citep{mcculloch2000bayesian,natarajan2000monte}. In a Bayesian setting, normalizing the first diagonal entry of the
covariance matrix requires a non-standard prior on the covariance
matrix. To avoid this challenge, some papers propose to estimate the unidentified
model and rescale $\Sigma$ by its diagonal $D$ as $D^{-\frac{1}{2}}\Sigma D^{-\frac{1}{2}}$
to obtain a correlation matrix \citep{mcculloch1994exact}, which may
have higher standard errors or result in diverging estimates.
 An alternative to address the scale invariance in the fixed coefficient
model is to specify that $\beta_{1}=1$ or $\|\beta\|_2=1$. In our paper, we propose a novel identification constraint on the  trace of the precision matrix, i.e., $\text{Tr}(\Sigma^{-1})=c$, inspired by \citet{burgette2012trace}, who propose a Bayesian approach that constructs a symmetric prior based on a trace constraint on the covariance matrix. 

\textbf{Shift Invariance.} The second identification issue is unique to the multinomial probit model of discrete choice. In the specification $Y_{ij} = \mathbf{1}\{Z_{ij}=\max_{j}Z_{ij}\}$ in \eqref{eq:multinomial-probit}, $Y_{i}$ is invariant under a universal shift of all entries of $Z_{i}$:
adding the same (observation-specific) constant the to latent utilities $Z_{ij}$ does not change
the implied choice behavior. Consequently, conditional on $X_{i}$, there can be multiple values of $(\theta,\Sigma)$
that give rise to the same distribution of $Y_{i}$. One common identification strategy is to leverage the existence of an ``outside option'', which is chosen when the latent utilities of all alternatives are \emph{negative}:
\begin{align}
\label{eq:multinomial-probit-identified}
Y_{ij} & =\begin{cases}
1 & \text{if }Z_{ij}=\max_{j}Z_{ij}\geq0\\
0 & \text{otherwise}
\end{cases}.
\end{align}
This approach is particularly convenient for choice datasets where a decision maker may choose to select none of the $m$ alternatives. For example, in the Expedia hotel search dataset in \cref{sec:expedia}, a customer can choose to not make a booking with any of the displayed results. When there is no natural default alternative, a common identification strategy \citep{train2009discrete} is to designate one of the $m$ alternatives, e.g., the first alternative, as the default option, and subtract its latent utility from all coordinates of $Z_i$. The transformed model generates the same distribution on $Y_{i}$ as the original model, but can now be identified in the same way as \eqref{eq:multinomial-probit-identified}. In this paper, we assume a level identification strategy has been adopted, and solve the maximum likelihood estimation problem \eqref{eq:mle} with respect to $(\beta, \Sigma)$ of the identified model. 

\subsection{The General EM Algorithm for Probit Estimation}
We now describe the general Expectation-Maximization approach that we adopt in this paper for probit estimation, following \citet{natarajan2000monte},  although our specific EM algorithm differs considerably in both the E (\cref{sec:E-step}) and the M (\cref{sec:M-step}) steps. The main reason for adopting an EM approach instead of a direct approach is to avoid explicitly simulating or sampling from TMVN. As we will see shortly, the EM approach we propose only requires computing the first two moments of TMVN, which is approximated efficiently and deterministically using EP.  
\begin{algorithm}[h!]
    \begin{algorithmic}

        \While{Not converged}\\
        \textbf{E Step}
          \State $S_i \gets Var(Z_i | Y_i, X_i, \beta, \Sigma)$
          \State $\mu_i \gets \mathbb{E}(Z_i | Y_i, X_i, \beta, \Sigma)$\\
           \textbf{M Step}

            \State $\beta \gets (\sum_{i}X_{i}^{T}\Sigma^{-1}X_{i})^{-1}\cdot(\sum_{i}X_{i}^{T}\Sigma^{-1}\mu_{i})$
          
          \State $\hat S \gets 
            \frac{1}{n} \sum_{i=1}^n \left( S_i + (\mu_i - X_i\beta) (\mu_i - X_i\beta)^T \right) $
         
          \State  $\Sigma \gets \arg \max_{s(\Sigma)=0}  -\log\det(\Sigma)-\text{Tr}(\Sigma^{-1}\hat{S})$
          
        \EndWhile
        \caption{Generic EM Algorithm for Probit Estimation}
    \label{alg:em_generic}\end{algorithmic}
\end{algorithm}

In order to circumvent challenges of directly optimizing complicated non-convex log-likelihoods such as that of the probit model in \eqref{eq:mle}, the EM algorithm originally proposed by \citet{dempster1977maximum} simply alternates between maximizing the expectation of the complete data log-likelihood \emph{conditional} on current parameter estimates, and computing the new conditional expectations based on updated parameters. For the probit model, the \emph{complete} data log-likelihood given latent variables $Z_1,\dots,Z_n$
is 
\begin{align*}
\ell(\beta,\Sigma\mid Z_{1},\dots,Z_{n}) & = -\frac{n}{2}\log\det(\Sigma)-\frac{1}{2}\text{Tr}(\Sigma^{-1}\cdot\sum_{i=1}^{n}(Z_{i}-X_{i}\beta)(Z_{i}-X_{i}\beta)^{T}).
\end{align*}
Computing the conditional expectation of the complete data log-likelihood above given current parameter estimates $\beta^{(t)},\Sigma^{(t)}$ results in the following lower bound of \eqref{eq:mle} (modulo $n/2$):
\begin{align}\label{eq:lb}
 \mathbb{E}(\ell\mid \{(X_{i},Y_{i})\}_{i=1}^{n},\beta^{(t)},\Sigma^{(t)}) \propto -\log\det(\Sigma)-\text{Tr}(\Sigma^{-1}\mathbb{E}(S\mid \{(X_{i},Y_{i})\}_{i=1}^{n},\beta^{(t)},\Sigma^{(t)})),
\end{align}
where $S$ is the sample covariance
\begin{align*}
  S = \frac{1}{n} \sum_{i=1}^n (Z_i - X_i\beta)(Z_i - X_i\beta)^T.
\end{align*}
The EM algorithm then obtains new estimates of $(\beta,\Sigma)$ by maximizing the lower bound in \eqref{eq:lb}.

The E (expectation) step rests on
calculating, for each observation $i$,
\begin{align}
\label{eq:e-step}
\mathbb{E}((Z_{i}-X_{i}\beta)(Z_{i}-X_{i}\beta)^{T}\mid X_{i}, Y_{i},\beta^{(t)},\Sigma^{(t)}) & =(\mu^{(t)}_{i}-X_{i}\beta+(Z_{i}-\mu^{(t)}_{i}))(\mu^{(t)}_{i}-X_{i}\beta+(Z_{i}-\mu^{(t)}_{i}))^{T}\\
 & =S_{i}^{(t)}+(\mu^{(t)}_{i}-X_{i}\beta)(\mu^{(t)}_{i}-X_{i}\beta)^{T},
\end{align}
where $\mu_{i}^{(t)}=\mathbb{E}(Z_{i}\mid X_i, Y_{i},\beta^{(t)},\Sigma^{(t)})$
is the conditional expectation of $Z_{i}$ given observation
$Y_{i}$ and $t$-th step estimates $\beta^{(t)},\Sigma^{(t)}$. Similarly, 
$S_i^{(t)}=Var(Z_{i}\mid X_i, Y_{i},\beta^{(t)},\Sigma^{(t)})$ is the conditional covariance matrix of $Z_i$. Given $X_i, Y_{i},\beta^{(t)},\Sigma^{(t)}$, we know that $Z_{i}$ follows a truncated multivariate normal (TMVN) distribution, so the E step reduces to computing the first two moments of the TMVN. 

In the M (maximization) step, we need to maximize the lower bound \eqref{eq:lb} by solving 
\begin{align}
\label{eq:m-step}
\max_{\beta,s(\Sigma)=0} - \log\det(\Sigma)-\text{Tr}(\Sigma^{-1}\hat{S}),
\end{align}
subject to an identification constraint $s(\Sigma)=0$, and the conditional sample covariance $\hat S= \mathbb{E}(S\mid \{(X_{i},Y_{i})\}_{i=1}^{n},\beta^{(t)},\Sigma^{(t)})$ is given by 
\begin{align*}
  \hat{S} = \frac{1}{n} \sum_{i=1}^n S_{i}^{(t)}+(\mu^{(t)}_{i}-X_{i}\beta)(\mu^{(t)}_{i}-X_{i}\beta)^{T}.
\end{align*}
The maximization in \eqref{eq:m-step} is usually done in two steps, as the joint optimization
problem in $(\beta,\Sigma)$ does not have a closed form solution, but the individual optimization problems with respect to $\beta$ and $\Sigma$ are both convex. First, using $\Sigma^{(t)}$
from the $t$-th step, take first order condition with respect to
$\beta$ to obtain 
\begin{align*}
\sum_{i}X_{i}^{T}(\Sigma^{(t)})^{-1}(\mu^{(t)}_{i}-X_{i}\beta^{(t+1)}) & =0,
\end{align*}
 which results in the generalized linear regression type update 
\begin{align*}
\beta^{(t+1)} & =(\sum_{i}X_{i}^{T}(\Sigma^{(t)})^{-1}X_{i})^{-1}\cdot(\sum_{i}X_{i}^{T}(\Sigma^{(t)})^{-1}\mu^{(t)}_{i}).
\end{align*}
 Then we can plug the updated $\beta^{(t+1)}$ into $\hat S$ and maximize \eqref{eq:m-step} with respect to $\Sigma$. This coordinate descent type procedure is proposed by \citet{meng1993maximum}, who show that the resulting EM procedure enjoys the same convergence guarantees as those of the original EM algorithm \citep{wu1983convergence}. 
 
 The generic EM approach for probit estimation is summarized in \cref{alg:em_generic}. Importantly, the computational efficiency of EM depends on how computations in the E and M steps are carried out. The Monte Carlo EM algorithm of \citet{natarajan2000monte} is a popular instance of the generic EM approach. They propose to use a Gibbs sampler to sample from TMVN and construct the conditional expectations \eqref{eq:e-step} in the E step, and use the constraint $\Sigma_{11}=c$ in the M step \eqref{eq:m-step}. Unfortunately, both choices result in computationally intensive procedures: Gibbs sampling is sequential and can experience slow-down on large-dimensional TMVNs, and $\Sigma_{11}=c$ results in a non-linear optimization problem. In this work, we propose to use expectation propagation (EP) to approximate the conditional expectations in the E step, and the identification constraint $\text{Tr}(\Sigma^{-1})=c$ in the M step. 

\section{Related Works}
\label{sec:literature-review}
In this section, we review existing works on probit estimation. Readers more familiar with the literature can skip to \cref{sec:EP-Newton-EM}. The binary probit estimation problem was studied as early as \citet{fisher1935case}. Over the years, many alternative proposals to probit estimation have been studied. Here we survey the literature according to two general approaches: frequentist methods and Bayesian methods.

\textbf{Frequentist Methods.} Most frequentist approaches center on the task of approximating
the intractable integrals associated with truncated multivariate normal densities. A prominent simulation-based approach builds on the simple idea that replaces integrals with average over samples drawn from TMVN. For example, the popular simulated maximum likelihood estimation method \citep{lerman1981use,hajivassiliou1994classical,hajivassiliou1996simulation} approximates the probit likelihood using samples drawn from TMVN: 
\begin{align*}
\int_{A_{i}} f(Z_i,X_i;\beta, \Sigma)dZ_{i} \approx \frac{1}{|L|} \sum_{\ell=1}^L f(Z_i^{(\ell)},X_i^{(\ell)};\beta, \Sigma),
\end{align*} 
where $Z_{i} =X_{i}\beta+\epsilon_{i}$ and $A_{i}$ is the appropriate truncated region of integration depending on
the probit model. The approximate log-likelihood can then maximized directly using a variety of optimization techniques including Newton-Raphson and BFGS. Simulated likelihood estimators are asymptotically equivalent to MLE when $L = \Omega(\sqrt{N})$, i.e., the number of draws from TMVN grows with the square root of the sample size. Other simulation based proposals with better theoretical properties include the method of simulated moments \citep{mcfadden1989method} and method of simulated scores \citep{hajivassiliou1998method}. In simulated moment, an approximation of the likelihood $L(\theta)$ is substituted into the objective for a moment estimator
\begin{align*}
\theta_{mm} & =\arg\min_{\theta}(d-L(\theta))^{T}W^{T}W(d-L(\theta)),
\end{align*}
 where $W$ is some given fixed distance matrix, for example derived
from the first order conditions evaluated at some fixed $\theta^{0}$,
and $d$ is an indicator of whether each alternative is chosen in
every observation. When $\theta^{0}=\theta^{\ast}$ or a consistent
estimator of $\theta^{\ast}$, this method yields an estimator that
is asymptotically equivalent to the MLE. \citet{mcfadden1989method} suggests several 
sampling methods to approximate the likelihood $L(\theta)$, and also
suggests a iterative approach where the $\theta$ in $W$ is regularly
updated by current estimates. 

To achieve efficient approximation of the likelihood, the most popular technique is the GHK sampler \citep{geweke_efficient_1991,hajivassiliou1996simulation,keane1994computationally}. The GHK sampler employs a separation of variables technique based on the Cholesky factorization of the covariance to reduce the problem to sequentially sampling from one-dimensional truncated normals. Other Monte Carlo methods such as the Gibbs sampler can also be used to draw from TMVN when approximating the likelihood. 

A popular alternative simulation-based method is the Monte Carlo EM algorithm \citep{wei1990monte,chib1998analysis,natarajan2000monte}. This
approach also employs Monte Carlo samples of TMVN, but instead of approximating the likelihood directly, it approximates the conditional expectation of the complete data log-likelihood based on current parameter estimates, which is then maximized to update the estimates. For example, \citet{natarajan2000monte} use a Gibbs
sampler that cycles through the components of the latent normal variable.
Approximating the conditional expectations in the E-step requires drawing a long chain of latent variables and then averaging over
samples drawn after some burn-in period. \citet{levine2001implementations} propose a Monte Carlo EM algorithm based on importance
sampling. Notably, they suggest recycling draws from previous iterations
 of the EM algorithm through importance sampling, in order to avoid drawing
an MC sample for each iteration of the EM algorithm. \citet{moffa2014sequential} propose another alternative to Gibbs
sampling using a sequential Monte Carlo sampling method that consists
of two stages. In the first stage samples are drawn from truncated
multivariate Student $t$-distributions, which are then updated towards
truncated multivariate normal. This allows the samples of the latent
variables to be recycled and updated from one iteration to the next
rather than re-sampled at each iteration. Other works on probit estimation, including \citet{mandt_sparse_2017} and \citet{karmakar_understanding_2021}, have also used the EM algorithm without resorting to simulations for the E step. In particular \citet{mandt_sparse_2017} also proposes to use the EP method to approximate the moments of TMVN. However, their probit model has a particular factor structure for the covariance matrix rather than using a general covariance matrix. Moreover, they do not use the symmetric trace constraint. 

\textbf{Bayesian Methods.} Bayesian analysis of probit models has been a prominent research area, since it has a natural conjugate prior. Bayesian methods for probit are often variations on the theme of iterating between
sampling latent data and estimating parameters based on posterior
distribution, often with parameter expansion and data augmentation
beyond sampling latent data to simplify the computation of posterior. For example, methods based on Gibbs sampling  \citep{mcculloch1994exact,chib1998analysis} iteratively alternate
between sampling latent variables using a Gibbs sampler and drawing
parameters from the posterior distribution. The recent work of \citet{loaiza2021scalable} uses MCMC sampling based on a symmetric prior on covariance matrices with a factor structure and achieve   scalability to larger choice sets. Their prior is constructed based on the symmetric trace constraint of \citet{burgette2012trace}. Such Bayesian methods differ from the Monte Carlo EM algorithm discussed earlier
in that after latent variables are drawn, they sample \emph{model parameters}
conditional on the latent variable using a prior on the parameters. A major challenge for the Bayesian approach to probit estimation is the selection of priors subject to identification constraints. \citet{mcculloch1994exact} ignore identification issues and use
the inverse Wishart distribution as the prior on covariance matrix $\Sigma$,
while \cite{chib1998analysis} restrict $\Sigma$ to be a \emph{correlation}
matrix, and uses a Metropolis-Hastings algorithm to sample the correlation
matrix with a uniform prior.

Parameter expansion with data augmentation \citep{liu1999parameter,imai2005bayesian,tabet2007bayesian} uses reparameterization
with data augmentation to aid in the computation of the posterior, and
requires $\Sigma$ to be a correlation matrix $R$. In the first step,
one draws the latent variable $Z$ conditional on observed variable
$Y$ and parameter $\theta=(\beta,R)$, i.e., from the TMVN,
which is achieved through Gibbs sampling from a series of univariate
truncated normals in the usual manner. In the second step, one augments
$\theta$ with $\alpha$ and draw $(\alpha,\theta)$ from the joint
posterior distribution. $\alpha$ is defined so that the expanded
model $p(Y,W\mid\theta,\alpha)$ with an appropriate prior yields
a posterior on $(\alpha,\theta)$ that is easy to compute through
another set of Gibbs sampling steps.

Works on Bayesian analysis of probit models have also studied the importance of identification conditions on the computation and statistical efficiency of the resulting estimators. For example, \citet{burgette2021symmetric} propose to address the shift invariance of multivariate probit models using the constraint that latent utilities sum to zero, instead of subtracting the latent utility of a ``reference alternative'' or ``base category'' from other coordinates. This is motivated by observations by \citet{burgette2012trace} that 
Bayesian multinomial probit predictions can be sensitive to the choice of the reference alternative, whose selection is often arbitrary and leads to an asymmetric prior. We draw inspirations from these works and use a symmetric identification constraint in the EM approach. 

It is also worth noting that EP and its close relative variational inference methods \citep{bleiVariationalInferenceReview2017} have been applied in Bayesian approaches to probit estimation. 
For example, \citet{fasano2023efficient} apply EP in Bayesian analysis of binary probit models with high-dimensional covariates. \citet{riihimaki2013nested} apply EP to approximate the posterior distribution of the covariate coefficient in probit models. However, the key difference to the present work is that in Bayesian approaches, EP is used for posterior approximation, whereas in our paper, EP is used instead for \emph{moment} approximation.

Most frequentist approaches
and Bayesian methods require efficient simulation or sampling from the truncated multivariate normal distribution. They
differ in how such samples are used. For example, direct simulation based methods such as simulated maximum likelihood use the samples to approximate the likelihood; the EM approach uses
these draws to approximate the conditional expectation of the complete data log-likelihood;
Bayesian methods use sampled draws to calculate the posteriors
of parameters. 

\textbf{What's New in Our Work.} Despite the theoretical and computational developments for probit estimation in the aforementioned works, the practical application of probit models has been limited by the significant computational challenges of simulation-based methods. 
In this paper, we circumvent the \emph{harder} problem of simulating or sampling from TMVN distributions by considering an alternative deterministic approach to directly approximate the moments of TMVN. 
\citet{minka2001expectation,minka2001family} proposed and developed the Expectation propagation (EP) framework, originally in the context of variational inference \citep{bleiVariationalInferenceReview2017}. 
Since its conception, EP has been predominantly applied in Bayesian analysis. \citet{minkaDivergenceMeasuresMessage2005} provides a detailed discussion of EP in relation to other methods for approximate Bayesian inference, such as message passing and variational Bayes. \citet{cunningham_gaussian_2013} used EP to approximate multivariate normal distributions truncated to linearly-constrained regions, specifically to estimate multivariate normal probabilities.  
In our work, we apply EP to approximate TMVN moments in an EM framework, and also propose customizations of the EP method for the multinomial probit by leveraging the sparsity and involution properties of the constraint matrix, which could provide further acceleration for discrete choice problems.
\citet{mandt_sparse_2017} previously applied EP to estimate (multivariate) probit linear mixed models for feature selection, where the covariance matrix has a simple structure as a weighted average between the identity matrix and a linear kernel of the data matrix.
In contrast, our method applies to both the multivariate and multinomial probit models, and we do not impose any constraints on the covariance matrix except for the identification constraint.
We give a more detailed discussion of EP in the context of probit estimation in \cref{sec:E-step}.

Another innovation of our work is the use of a symmetric trace identification constraint on the precision matrix, which significantly simplifies the constrained optimization problem in the EM algorithm. This proposal is an interesting combination of ideas from Bayesian and frequentist approaches to probit estimation. \citet{burgette2012trace} proposed a symmetric trace identification constraint on the covariance matrix in a Bayesian framework. In our work, we place the constraint instead on the precision matrix, which results in considerable simplifications of the constrained optimization problem in the maximization step of our EM approach. In contrast, previous EM approaches to probit estimation often use asymmetric identification conditions on the covariance matrix, which result in a constrained nonlinear optimization problem that requires the use of optimization packages.
\section{Efficient Probit Estimation via EP-Newton EM Algorithm}
\label{sec:EP-Newton-EM}
In this section, we propose the EP-Newton EM algorithm, a new variant of the generic EM algorithm designed specifically for probit estimation. Compared to previous proposals of the EM algorithm for probit estimation, our algorithm is deterministic and does not require sampling from truncated multivariate normal distributions, and solves a one-dimensional optimization problem instead of large-dimensional nonlinear programs. 

\subsection{E Step: Expectation Propagation for Truncated Multivariate Normal Moments}
\label{sec:E-step}
It is well-known that computing the truncated multivariate normal (TMVN) moments \eqref{eq:e-step} is expensive for large choice sets in the multivariate probit model. 
The expectation propagation (EP) approximation of \citet{cunningham_gaussian_2013} can provide a fast and relatively accurate estimate of the moments. 
The idea is to approximate the target TMVN density with an unrestricted multivariate normal density using EP. 
We use the mean and covariance of the approximating multivariate normal distribution as estimates of $\mathbb{E}[Z_i|Y_i,X_i,\beta, \Sigma]$ and $Var[Z_i|Y_i,X_i,\beta, \Sigma]$ in the E step of \cref{alg:em_generic}, which is the first distinction of this work compared to previous methods. 
We next provide a brief discussion of the EP method, and direct the readers to \cref{sec:EP-detailed-intro} for a self-contained introduction to EP. 

\begin{figure}[h!]
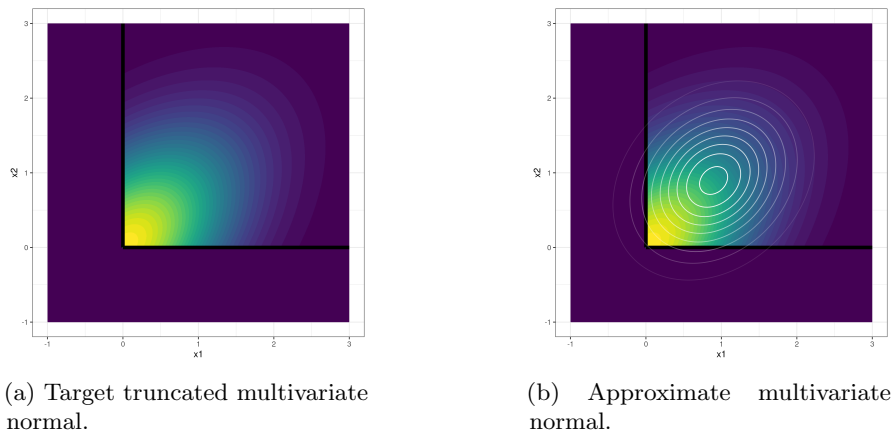

    \centering
    \begin{subfigure}[b]{0.3\textwidth}
        \centering
        \includegraphics[width=\textwidth]{ep_exposition/target_tmvn.pdf}
        \caption{Target truncated multivariate normal.}
    \end{subfigure}\hspace{2cm}
    \begin{subfigure}[b]{0.3\textwidth}
        \centering
        \includegraphics[width=\textwidth]{ep_exposition/approx_mvn.pdf}
        \caption{Approximate multivariate normal.}
    \end{subfigure}
    \caption{EP multivariate normal approximation to positive orthant truncated multivariate normal.}
    \label{fig:ep_orthant_tmvn}
\end{figure}

The method proposed in \citeauthor{cunningham_gaussian_2013} applies variational inference to compute truncated multivariate normal density integrals. 
Given the task of approximating an unknown (posterior) distribution $p(z|x)$, the basic idea of variational inference is to find a density $q^*(z)$ from a tractable family $\mathcal{Q}$ that best approximates $p(z|x)$, by minimizing a divergence measure $D$ between the true posterior $p(z|x)$ and approximation $q(z) \in \mathcal{Q}$, where $\mathcal{Q}$ is a family of distributions over $z$. 
The approximating family $\mathcal{Q}$ needs to trade off flexibility, which allows accurate approximation, with computational speed. 
The choice of divergence $D$ also has an impact on the approximation. 
Gaussian expectation propagation \citep{minka2001expectation} sets $\mathcal{Q}$ to be the family of multivariate normal distributions, and minimizes a \emph{local} KL divergence $D$. 
This combination of choices leads to an approximation that captures the zeroth (normalizing constant), first, and second moments of the target posterior, which is exactly what is needed for truncated multivariate normal distribution in probit estimation. 
\citeauthor{cunningham_gaussian_2013} use EP to approximate truncated multivariate normals with unconstrained multivariate normal distributions (Figure~\ref{fig:ep_orthant_tmvn}), but the application of EP in the estimation of probit models is uncommon.

Now we discuss in detail the implementation of the EP method for truncated multivariate normal estimation in the EM algorithm for probit estimation. 
Importantly, in probit models the region of integration $\mathcal{A}_i$ is a subspace with linear constraints, i.e., a polytope. 
For the multivariate probit, $\mathcal{A}_i \subseteq \mathbb{R}^{m}$ is an axis-aligned rectangular region. 
For the multinomial probit model using the reference/default item identification strategy, we have $\mathcal{A}_i=\{x:l\leq A_i x \leq u\} \subseteq \mathbb{R}^{m-1}$, with 
\setcounter{MaxMatrixCols}{20}
\begin{align}\label{eq:Alongform}
{
  A_i: = 
    \begin{pmatrix}
      -1     &     0  & \cdots & \cdots & 0  & 1      & 0  & \cdots &          & \cdots & 0
      \\
      0      &     -1 & 0      & \cdots & 0  & 1      & 0  & \cdots &          & \cdots & 0
      \\
      \vdots &        & \ddots & \ddots &    & \vdots &    &        &          &        & \vdots   
      \\
      0      & \cdots &        &      0 & -1 & 1      & 0  & \cdots &          & \cdots & 0
      \\
      0      & \cdots &        & \cdots &  0 & 1      & 0  & \cdots &          & \cdots & 0
      \\
      0      & \cdots &        & \cdots &  0 & 1      & -1 & 0      &          & \cdots & 0
      \\
      \vdots &        &        &        &    & \vdots &    & \ddots & \ddots   &        & \vdots
      \\
      0      & \cdots &        & \cdots &  0 & 1      & 0  & \cdots & 0        & -1     & 0
      \\
      0      & \cdots &        & \cdots &  0 & 1      & 0  & \cdots &          & 0      & -1 
    \end{pmatrix}
    }
\end{align}
when the chosen item is not the reference alternative, 
where the row (and column) of $A_i \in \mathbb{R}^{(m-1)\times(m-1)}$ corresponding to the selected non-reference alternative is distinct from the other rows. 
This non-reference alternative constraint matrix can also be written as 
\begin{align}\label{eq:Ashortform}
    A_i = -I_{m-1} + \mathbf{1}_{m-1}e_j^T + e_je_j^T,    
\end{align}
where $I_{m-1}$ is the identity matrix of dimension $m-1$, $\mathbf{1}_{m-1}$ is the vector of ones of dimension $m-1$, and $e_j$ is the $j$th basis vector of dimension $m-1$.
When the reference or default alternative is selected, $A_i$ is simply the identity matrix $I_{m-1}$. In both cases, the lower and upper bounds are given by
\begin{align}
  l = 
    \begin{pmatrix}
      0 \\ \vdots \\ 0
    \end{pmatrix},
  \qquad
  u = 
    \begin{pmatrix}
      \infty \\ \vdots \\ \infty
    \end{pmatrix}.
\end{align}
In our proposed algorithm, we apply the EP method to approximate moments of the truncated normal distributions on $\mathcal{A}_i$. When $\mathcal{A}_i$ is an axis-aligned region, EP exhibits faster convergence \citep{cunningham_gaussian_2013}. This is the case for the multivariate probit; for the multinomial probit of discrete choice, $\mathcal{A}_i$ is not axis-aligned, but note that $A_i$ is an \emph{involutory} matrix, i.e., $A_i^{-1} = A_i$. This property could potentially be useful in helping researchers to further customize and accelerate the EP method for probit estimation. We discuss this proposal in detail in Appendix \ref{sec:faster_ep}. For other specifications of the probit model such as the one in \citet{hausman1987specifying}, we can also apply EP to approximate the conditional moments, using potentially different $A_i$ and $l,u$.

\subsection{M Step: Trace Constrained Optimization via Newton's Method}
\label{sec:M-step}
In this section, we discuss the second contribution of our proposed EM algorithm for probit estimation.
In simulation studies, we observe that nonlinear optimization with respect to $\Sigma$ of the conditional likelihood \eqref{eq:m-step} in the M step is numerically unstable, which prevents the solution of problems with dimensions larger than 10. 
 This challenge motivates us to propose to solve a constrained optimization problem with respect to the \emph{precision} matrix $\Omega=\Sigma^{-1}$ in the M step, which is convex. Moreover, instead of using the asymmetric constraint $\Sigma_{11}=1$, we impose a symmetric trace constraint on $\Omega$ instead of $\Sigma$ to ensure identification, leading to the following version of the optimization problem \eqref{eq:m-step}:
\begin{align}
\label{eq:precision-trace-normalized}
\min_{\text{Tr}(\Omega)=m}-\log\det(\Omega)+\text{Tr}(\Omega\cdot \hat{S}).
\end{align}
Our trace constrained optimization approach differs from that in \citet{burgette2012trace}, which imposes the constraint on the covariance matrix $\Sigma$ and proposes a prior that enforces this constraint. In contrast, we enforce the constraint directly in the optimization problem in the M step of \cref{alg:em_generic}. It has been observed in practice that a trace normalization is usually preferred to fixing a diagonal entry of $\Omega$ or $\Sigma$. For example, suppose the true precision matrix $\Omega$ satisfies both $\text{Tr}(\Omega) = m$ and $\Omega_{11} = 1$. In numerical studies,
the former constraint on the trace usually performs much better than the latter, because it constrains
the entire diagonal and is thus a more uniform constraint. It will be interesting to further study settings where the trace constraint is superior to other asymmetric constraints. As we now demonstrate, the symmetric constraint has simple yet powerful implications for the optimization of \eqref{eq:precision-trace-normalized}. 

The constrained optimization problem in \eqref{eq:precision-trace-normalized} is a trace constrained version of the standard maximum likelihood estimation of the covariance matrix, where $\hat S$ is the conditional version of the empirical covariance matrix. With $y\in\mathbb{R}$ the optimal dual variable, the constrained optimization problem \eqref{eq:precision-trace-normalized} has optimal conditions
\begin{align}
\Omega=(\hat S-yI)^{-1}\succ0,\quad & \text{Tr}((\hat S-yI)^{-1})=m.
\end{align}
Note that $\text{Tr}((\hat S-yI)^{-1})$ is monotonically increasing in $y\in(-\infty,\lambda_{1})$, where $\lambda_{1}$ is
the smallest eigenvalue of $S$. This suggests that we can use bisection to find the root of $Tr((\hat S-yI)^{-1})=m$. More precisely, we can start with a lower bound $\underline{y}$ and upper bound $\overline{y}$ such that $Tr((\hat S-\underline{y}I)^{-1})<m$ and $Tr((\hat S-\overline{y}I)^{-1})>m$. If their midpoint satisfies $Tr((\hat S-\frac{\underline{y}+\overline{y}}{2}I)^{-1})<m$, then we know that $y\in (\frac{\underline{y}+\overline{y}}{2},\overline{y})$, so we update the new lower bound $\underline{y} \leftarrow \frac{\underline{y}+\overline{y}}{2}$. Otherwise, update the new upper bound $\overline{y} \leftarrow \frac{\underline{y}+\overline{y}}{2}$. 

Although the bisection procedure works effectively in practice to solve \eqref{eq:precision-trace-normalized} in place of off-the-shelf optimization solvers, we can improve upon it even further. Let  $\lambda_{1}\leq\dots\leq\lambda_{m}$
be the eigenvalues of $\hat S$ in ascending order. The optimality condition then becomes the equation
\begin{align}
\label{eq:Newton-root}
 \sum_{i}\frac{1}{\lambda_{i}-y}=m, \quad y<\lambda_{1}.
\end{align}
As $f(y):=\sum_{i}\frac{1}{\lambda_{i}-y}-m$ is an increasing function for $y\in(-\infty,\lambda_{1})$, its root can be found efficiently using Newton's method: 
\[ y^{k+1} = y^k- {f(y^k)}/{f'(y^k)}.\]
Both bisection and Newton's method are guaranteed
to find the optimal $\Omega=(\hat S-yI)^{-1}$, and we recommend using a hybrid approach that starts with a bisection procedure to locate a starting point to the right of the optimal $y$, and then proceed using Newton's method \citep{ye1992new}. Instead of relying on optimization packages, in our code we implement a simple customized solver based on the Newton method, which is able to achieve significant speedup over off-the-shelf convex optimization packages such as \texttt{CVX}. As a result, for large problems $(m > 100)$, our customized solver can obtain accurate estimators of the covariance matrix, while previous methods using package solvers fail to converge in reasonable time. We leave a formal study of the acceleration effect of the symmetric constraint on the optimization landscape of \eqref{eq:m-step} in the M step to future works. 

Lastly, recall that the E step in the EM algorithm requires the updated covariance matrix matrix $\Sigma$ instead of the precision matrix, so in the M step, we can actually avoid taking the inverse by directly setting 
\[\Sigma^{(t+1)}=\hat S-y^\ast I,\] 
where $y^\ast$ is the solution to \eqref{eq:Newton-root}. We have therefore obtained the customized EM algorithm based on EP and Newton's method, which we describe in \cref{alg:em_customized}. 
\begin{algorithm}[h!]
    \begin{algorithmic}

        \While{Not converged}\\
        \textbf{E Step}\\ 
        Use EP to approximate 
          \State $S_i \gets Var(Z_i | Y_i, X_i, \beta, \Sigma)$
          \State $\mu_i \gets \mathbb{E}[Z_i | Y_i, X_i, \beta, \Sigma]$\\
           \textbf{M Step}

            \State $\beta \gets (\sum_{i}X_{i}^{T}\Sigma^{-1}X_{i})^{-1}\cdot(\sum_{i}X_{i}^{T}\Sigma^{-1}\mu_{i})$
          
          \State $\hat S \gets 
            \frac{1}{n} \sum_{i=1}^n \left( S_i + (\mu_i - X_i\beta) (\mu_i - X_i\beta)^T \right) $

           \State $\lambda_{1}\leq\dots\leq\lambda_{m} \gets $ eigenvalues of $\hat{S}$ 
         
          \State  $\Sigma \gets \hat S-y^\ast I,\quad y^\ast <\lambda_1, \sum_{i}\frac{1}{\lambda_{i}-y^\ast}=m$
          
        \EndWhile
        \caption{Our Proposed EP-Newton EM Algorithm for Probit Estimation}
    \label{alg:em_customized}\end{algorithmic}
\end{algorithm}

In the absence of a natural default alternative, the choice of the reference alternative to ensure identification is somewhat arbitrary \citep{train2009discrete}. Previous works have also observed that this choice can lead to different behaviors of the estimated model. To address this issue, \citet{burgette2021symmetric} propose a symmetric identification strategy. Instead of selecting a reference item, they propose to restrict the sum of latent utilities over all alternatives to be normalized to zero. This specification leads to a rank-deficient random utility model, and \citet{burgette2021symmetric} propose a Bayesian approach with a novel prior on the space of rank-deficient covariance matrices. This approach is further studied by \citet{pan_globally_2023}. 
 See also \citet{cong2017fast}, which can be used to sample from covariance matrix restricted to planes. In our EM framework, such a symmetric identification strategy requires extending the EP algorithm to approximate rank-deficient truncated multivariate normal distributions, which we leave to future works.

\section{Simulation Studies}
\label{sec:simulations}
In this section, we conduct simulations to study the performance of our proposed EM algorithm compared to previous methods. In particular, we compare against the popular simulated maximum likelihood estimator using the GHK sampler \citep{geweke_efficient_1991,hajivassiliou1996simulation,keane1994computationally}, the MC-EM algorithm using Gibbs sampler \citep{natarajan2000monte}, and a customized MC-EM algorithm using Hamiltonian Monte Carlo sampler \citep{neal2011mcmc}. Our simulation experiments have $n=2000$ observations, $\beta=\mathbf{1}_p$ where $p=10$, and true covariance
   \[ \Sigma = 0.5 \boldsymbol{I}_m + 0.5 \boldsymbol{1}_m \boldsymbol{1}_m^T.\]
Data is generated according to
\begin{align*}
    X_i &\sim Unif_{m\times p}(-.5, .5)
    \\
    Z_i &= X_i\beta_i + \epsilon_i
    \\
    Y_i &= \max_j Z_{ij}
    \\
    \epsilon_i &\sim MVN(0_m, \Sigma).
\end{align*}
To ensure identification, we reduce the system by subtracting a reference utility to address shift invariance.

\subsection{Comparison with Simulated Likelihood on Small Problems}
\begin{figure}[h!]
        \centering
        \begin{subfigure}[b]{0.3\textwidth}
            \centering
            \includegraphics[width=\textwidth]{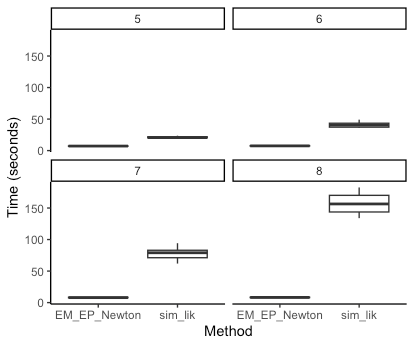}
            \caption{Runtime}
        \end{subfigure}
        \hfill
        \begin{subfigure}[b]{0.3\textwidth}
            \centering
            \includegraphics[width=\textwidth]{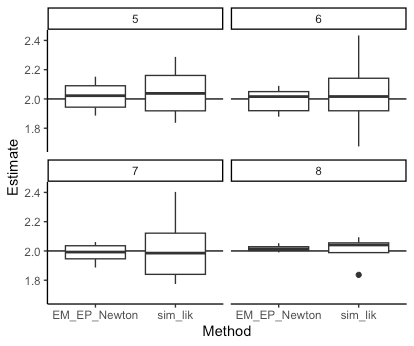}    
            \caption{$\beta$ Error}
        \end{subfigure}
        \begin{subfigure}[b]{0.3\textwidth}
            \centering
            \includegraphics[width=\textwidth]{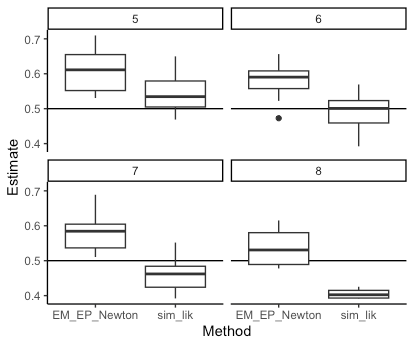}    
            \caption{$\Sigma$ Error}
        \end{subfigure}
        \caption{Comparison of runtime (left) and estimation error (right) of proposed  EM algorithm with simulated maximum likelihood.}\label{fig:comparison_small}
    \end{figure}

\cref{fig:comparison_small} contains the comparison of our method with simulated maximum likelihood, as implemented in the \texttt{R} package \texttt{mlogit}, on multinomial probit models with small choice sets. We see that the proposed EM algorithm has comparable accuracy when estimating $\beta,\Sigma$, but has considerably lower runtime. 

\subsection{Performance on Larger Problems}
\begin{figure}[h!]
        \centering
        \begin{subfigure}[b]{0.35\textwidth}
            \centering
            \includegraphics[width=\textwidth]{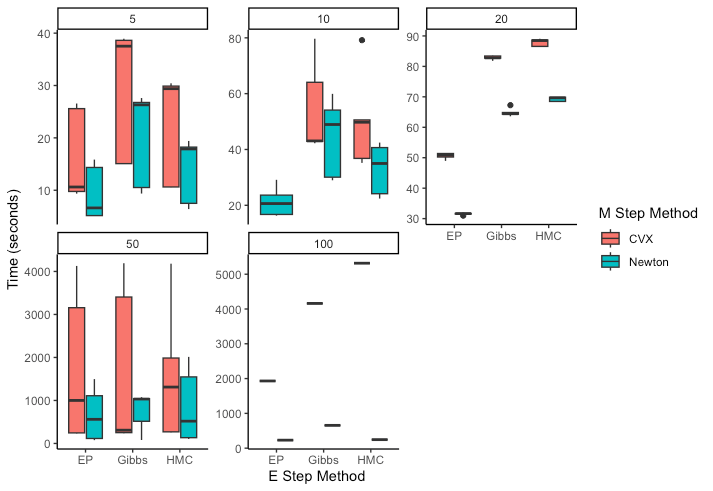}
            \caption{Runtime}
        \end{subfigure}
        \hfill
        \begin{subfigure}[b]{0.3\textwidth}
            \centering
            \includegraphics[width=\textwidth]{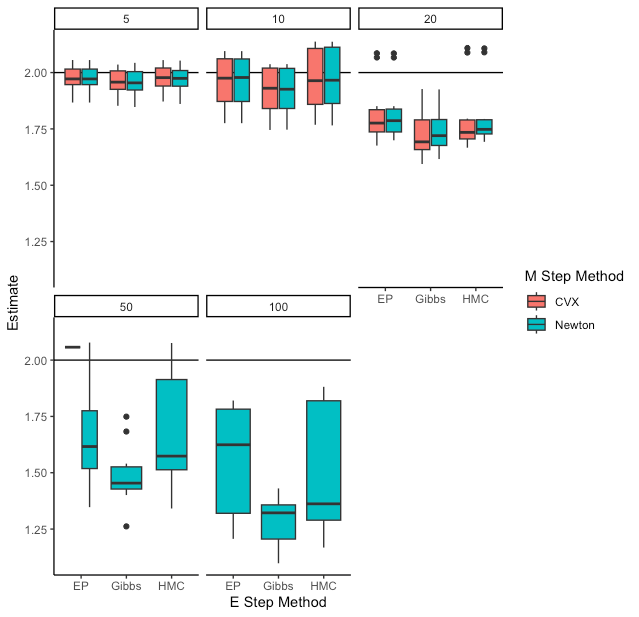}    
            \caption{$\beta$ Error}
        \end{subfigure}
        \hfill
        \begin{subfigure}[b]{0.3\textwidth}
            \centering
            \includegraphics[width=\textwidth]{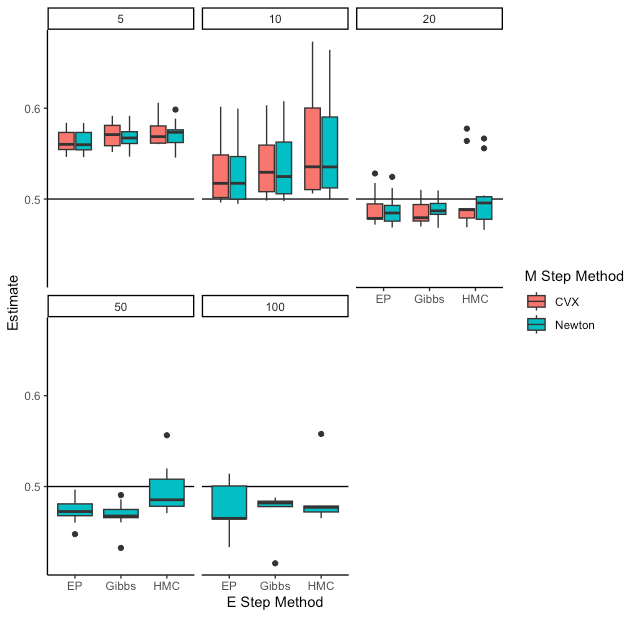}    
            \caption{$\Sigma$ Error}
        \end{subfigure}
        \caption{Comparison of runtime (left) and estimation error (right) of proposed  EM algorithm with MC-EM algorithms using Gibbs sampler and Hamiltonian MC sampler. Teal bars correspond to results based on solving the M step problem with \texttt{CVX} package, wihle red bars correspond to results based on solving the M step problem using our proposed Newton method.}\label{fig:comparison_large}
    \end{figure}

\cref{fig:comparison_large} contains comparisons of our method with two variants of the MC-EM algorithm based on the Gibbs sampler and the Hamiltonian MC sampler. Moreover, we compare the performance of our Newton method to the \texttt{CVX} package when solving the M-step optimization problem subject to the trace constraint. There are several messages. First, we see that Newton's method improves the computation time significantly over \texttt{CVX}, and for problem size larger than 50, \texttt{CVX} becomes computationally infeasible. Second, our proposed EM algorithm consistently outperforms the MC-EM algorithms in terms of both runtime and estimation accuracy. Third,  larger probit models (larger $m$) require more samples to achieve consistent estimation. The sample size of 2000 seem to be inadequate for $m>50$. Lastly, the MC-EM algorithm based on Hamiltonian MC sampler performs quite well, and it may be interesting to study further the complexities of different methods as well as their accuracies as $m$ increases.

\subsection{Comparison of Symmetric and Asymmetric Identification Constraints}

We also explore how different identification strategies affect the performance of probit estimation methods. More precisely, we compare the strategy that sets $\Sigma_{11}=1$ with our trace constraint $\text{Tr}(\Sigma^{-1})=m$ in our proposed EM algorithm.
With the former asymmetric specification, the M step needs to be solved using a non-linear solver, whereas with the trace constraint the M step is solved using the Newton method. For both specifications, the E step uses EP approximations. We find that for a small response variable $(m=6)$, trace identification has better runtime and accuracy for both the coefficient and the precision matrix, which suggests that the trace identification could be preferable in practice.

        \begin{figure}[h!]
        \centering
        \begin{subfigure}[b]{0.3\textwidth}
            \centering
            \includegraphics[width=\textwidth]{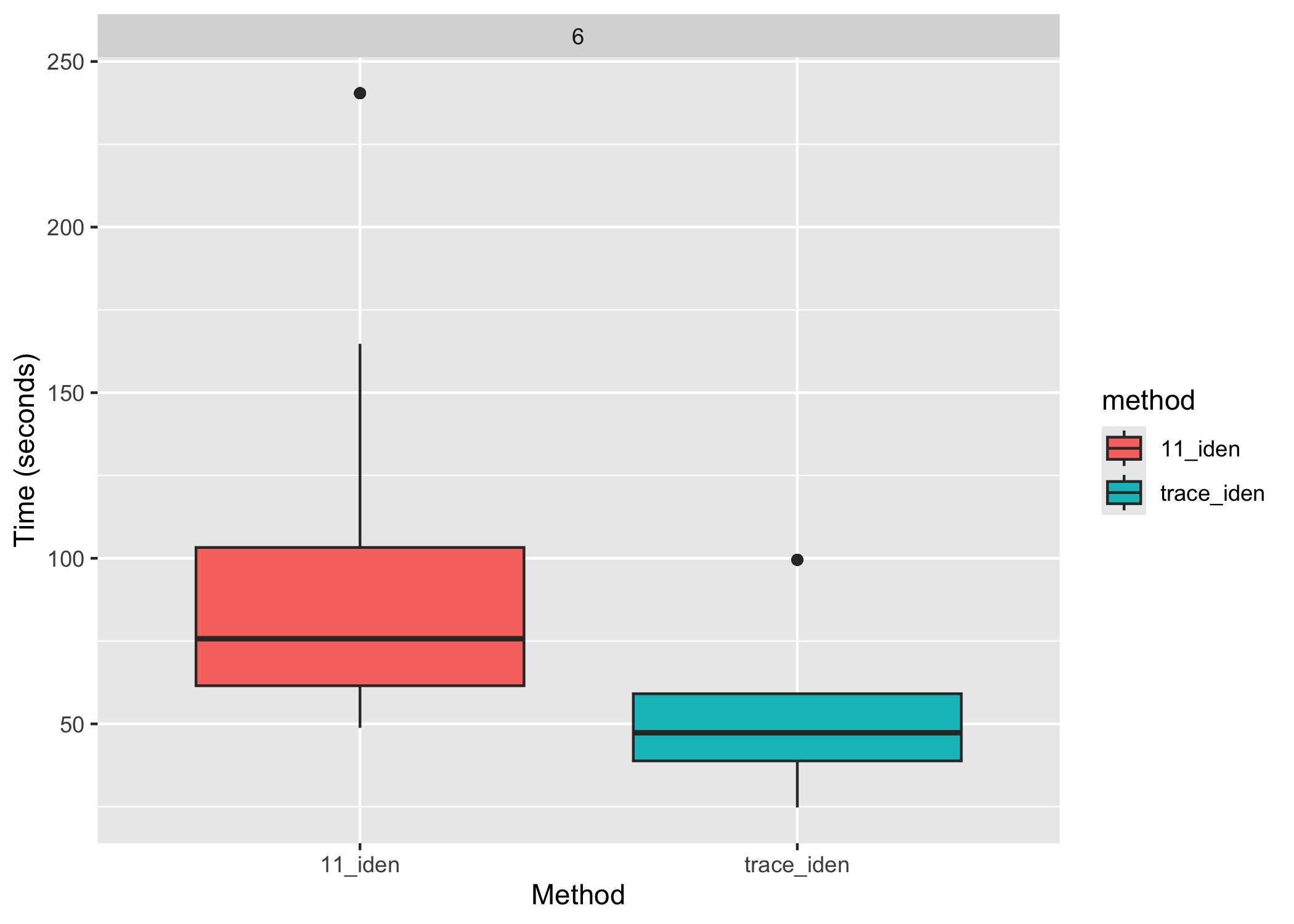}
            \caption{Runtime}
        \end{subfigure}
        \hfill
        \begin{subfigure}[b]{0.3\textwidth}
            \centering
            \includegraphics[width=\textwidth]{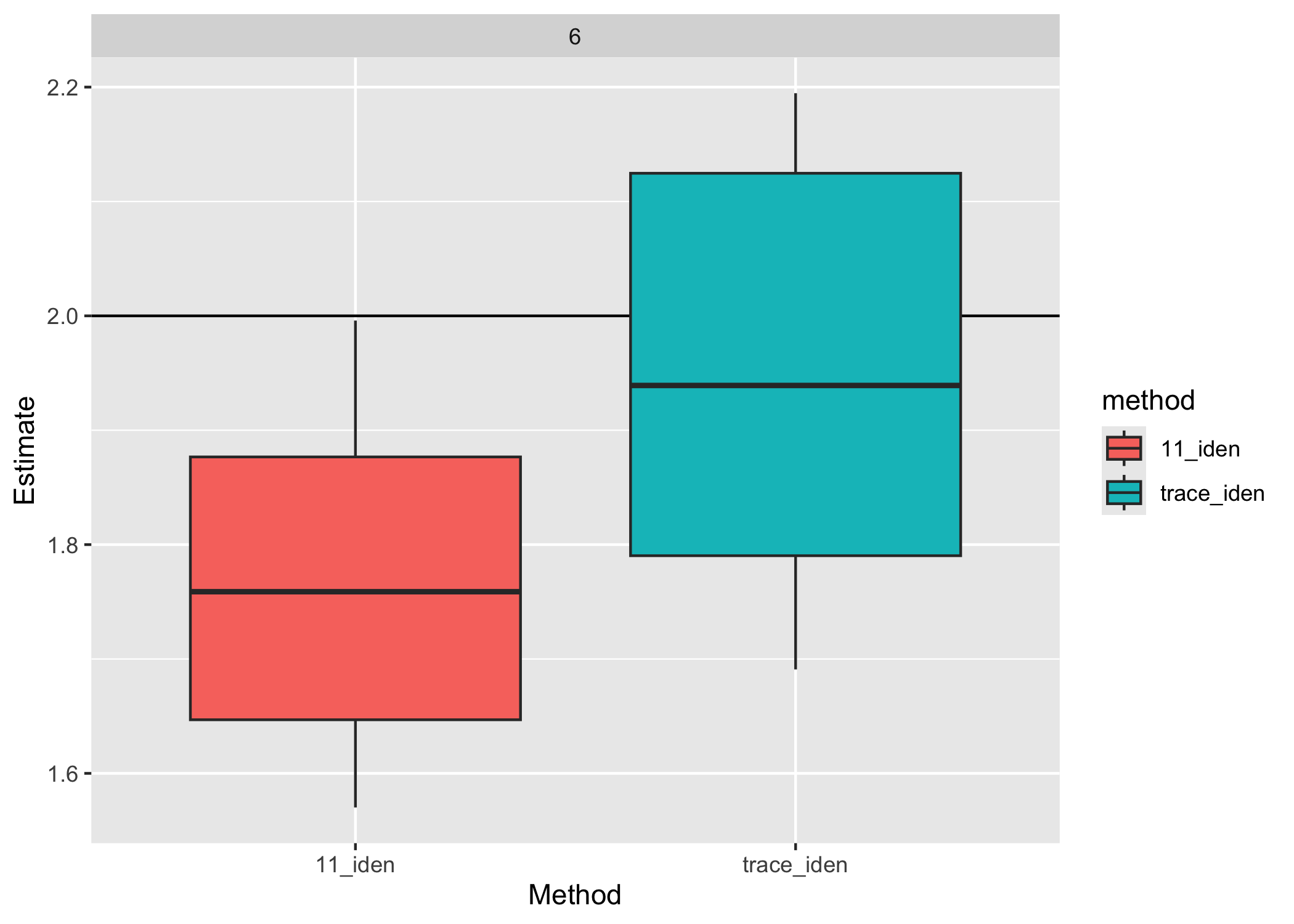}    
            \caption{$\beta$ Error}
        \end{subfigure}
        \hfill
        \begin{subfigure}[b]{0.3\textwidth}
            \centering
            \includegraphics[width=\textwidth]{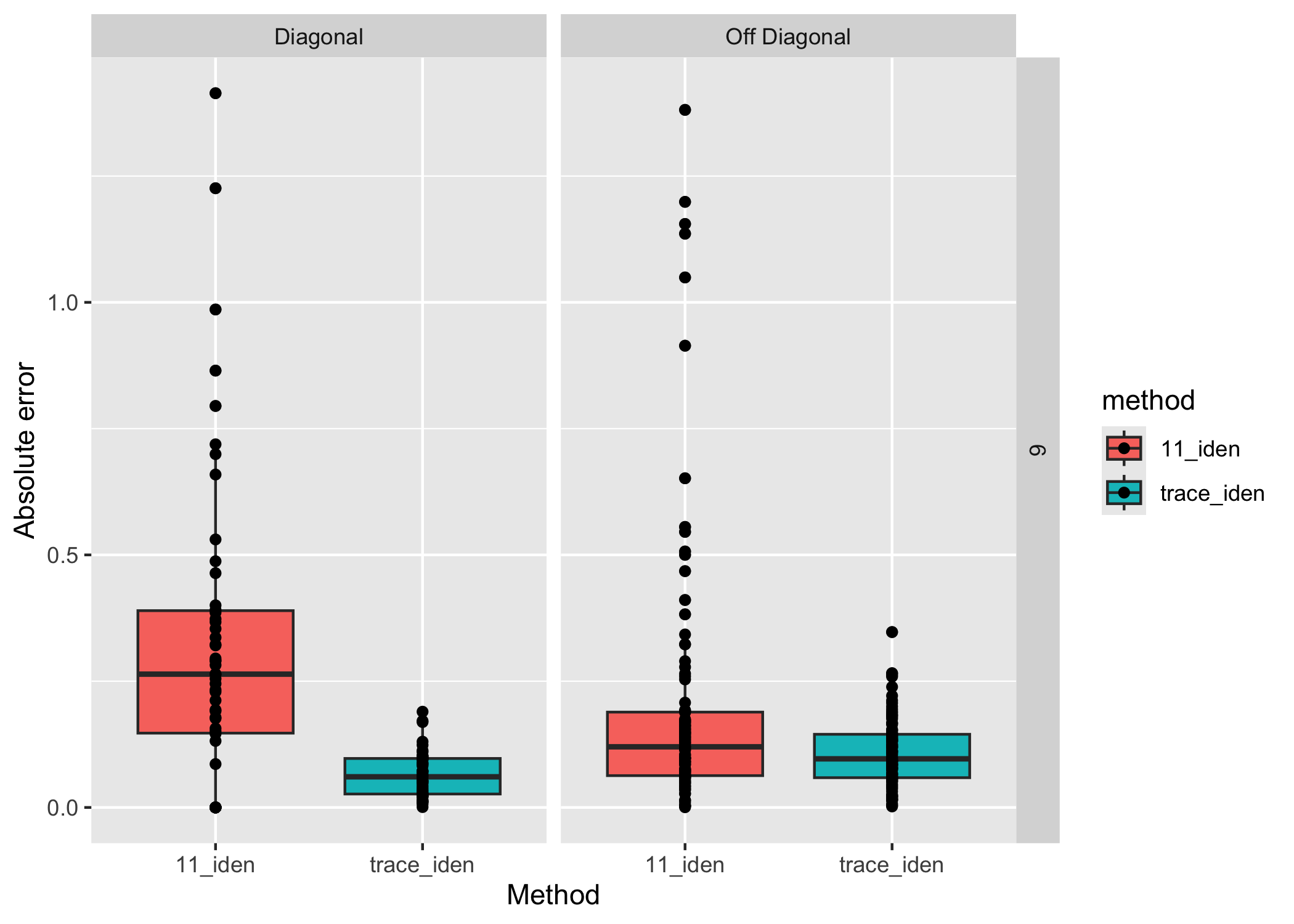}    
            \caption{$\Sigma^{-1}$ Error}
        \end{subfigure}
        \caption{Comparison of asymmetric and trace restrictions for precision matrix identification. Teal bars correspond to results based on trace restriction, while red bars correspond to results based on fixing $1,1$ element of precision to 1.}\label{fig:comparison_identification}
    \end{figure}

    \section{Application to Expedia Hotel Search Data}
    \label{sec:expedia}
    \begin{figure}[t]
        \centering
        \begin{subfigure}[b]{0.4\textwidth}
            \centering
            \includegraphics[width=.8\textwidth]{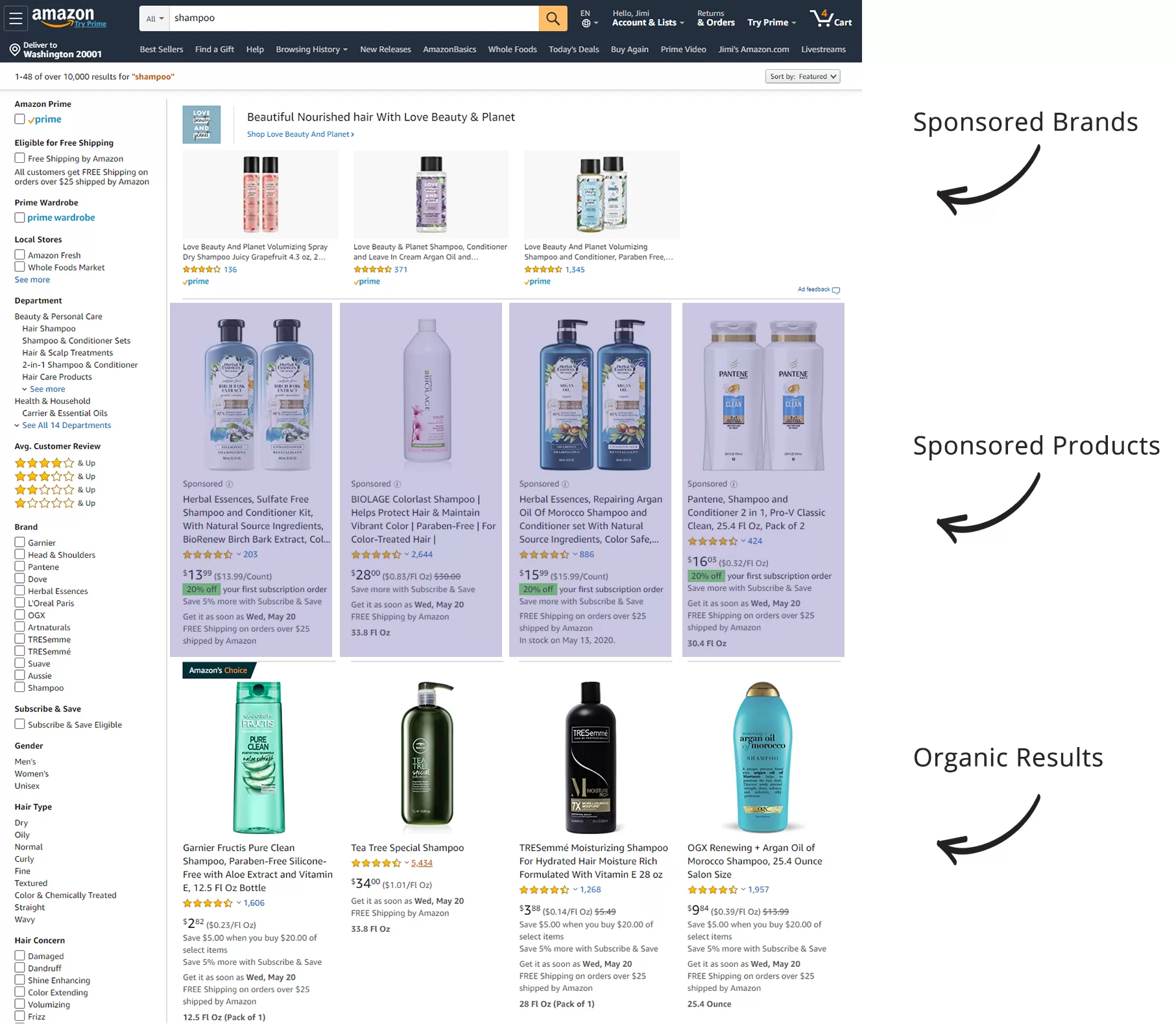}
        \end{subfigure}  
        \begin{subfigure}[b]{0.308\textwidth}
            \centering
            \includegraphics[width=.8\textwidth]{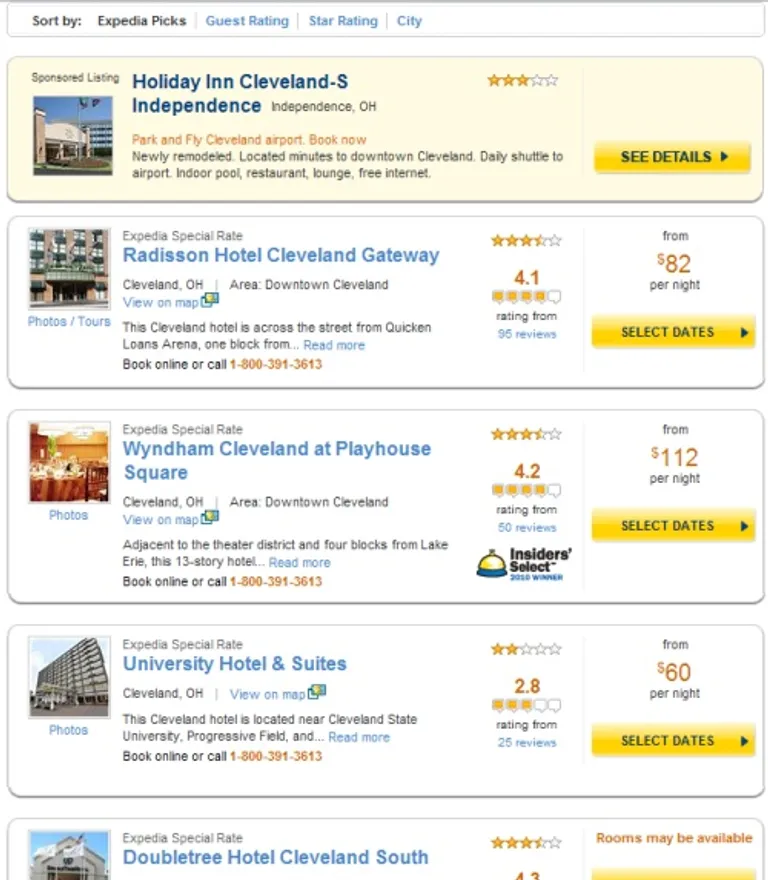}
        \end{subfigure}
        \caption{Sponsored listings in Amazon.com and Expedia.com search results.}
        \label{fig:sponsored}
    \end{figure}
 In this section, we demonstrate how one can use probit models to estimate contextual effects that may not be adequately captured by logit type models, leveraging ranked hotel search data from Expedia.com. 
\subsection{Motivating Question}
In online shopping and booking platforms, search results are often displayed as a ranked list or a panel, and the ordering of alternatives in a choice set can have significant impacts on consumers' choice behavior  \citep{ursu2018power}. Many platforms offer sellers and merchants the option of paying for ``sponsored'' search results that are displayed above regular search results, which can help merchants boost their sales or bookings \citep{long2022designing,gerpott2022competitive,joo2024sellers}. \cref{fig:sponsored} shows examples of sponsored listings in Amazon.com and on Expedia.com search results.

An important problem for the platform is the pricing of such sponsored slots. While the simplest solution is to offer a single price to all sellers who are interested in promoting their products, a more sophisticated and individualized pricing plan can help the platform better capture potential revenue. More precisely, the platform can ask the following counterfactual question: 
\begin{quote}
    Given a ranked list of search results output by the platform's baseline algorithm, what is the average  increase in demand (purchase probability) of the $k$-th item if it was instead placed on \emph{top} of the list, holding all other items' relative positions unchanged?
\end{quote}
Importantly, this increase in purchase probability could display considerable heterogeneity, both in the sponsored product itself, as well as in the \emph{context} of alternative options displayed in the search result. A higher increase justifies a higher charge by the platform to place the item in a ``sponsored'' slot. Therefore, the resulting problem can be understood as that of studying the \emph{positional} effects of a ranked list of alternatives, while holding the items in a bundle fixed. Although pricing in the presence of contextual effects has been a well-studied topic \citep{gallego2014multiproduct,bayer2020impact}, there has been relatively less attention on the problem of pricing in the context of \emph{ranked} consideration sets.

Note that in this problem, the ordered structure of the data is in the choice set instead of the \emph{response variable}, for which other models such as the ordered logit \citep{mccullagh1980regression} and the exploded logit \citep{plackett1975analysis,beggs1981assessing} can be employed. In contrast, when the choice set or bundle has an ordered structure, other approaches are needed for modeling the resulting choice data. A naive approach based on logit models is to incorporate the ordering information using dummy variables of position, which results in a high-dimensional covariate vector. A generalized logit model for ordered alternatives was proposed by \citet{small1987discrete}, while other approaches such as consideration set models based on limited attention or searching costs can also be used to study the demand increase arising from reordering \citep{caplin2019rational,abaluck2021consumers}.

In this paper, we argue that the probit model is in fact a natural choice to capture contextual effects in ordered choice sets, because the coordinates of a probit model's latent utility variables can be associated with the orders of items in a ranked bundle. Moreover, the outcome of no selection or purchase serves as a natural ``reference item'' for the purpose of identification instead of having to select an arbitrary reference coordinate. We can therefore estimate a standard probit model with a general covariance matrix on choice data with ranked bundles, as long as the data contains enough randomization with respect to the ordering of displayed items. The estimated model can be used to predict the purchase probabilities of specific items at any position of a displayed bundle. Moreover, there may be additional information on the substitution pattern in the estimated covariance or precision matrix.  

As data from such applications often include tens or hundreds of search results for a single search, a main challenge for using probit models is the large dimension of the response variable. Our proposed probit estimation method is therefore useful for estimating the ordering effects in discrete choice data with ranked choice sets. In this section, we will estimate a probit model using data from Expedia consisting of ``search result impressions'' \citep{Hamner_Friedman_Expedia_2013}, which are ordered lists of hotels that are displayed to a user after they input a search inquiry. 
Using the estimated probit model, we study the problem of pricing sponsored spots in online booking and shopping platforms with ordered displaying of options. 
\subsection{The Expedia Hotel Search Dataset}
\begin{figure}
        \centering
        \begin{subfigure}[b]{0.4\textwidth}
            \centering
            \includegraphics[width=.8\textwidth]{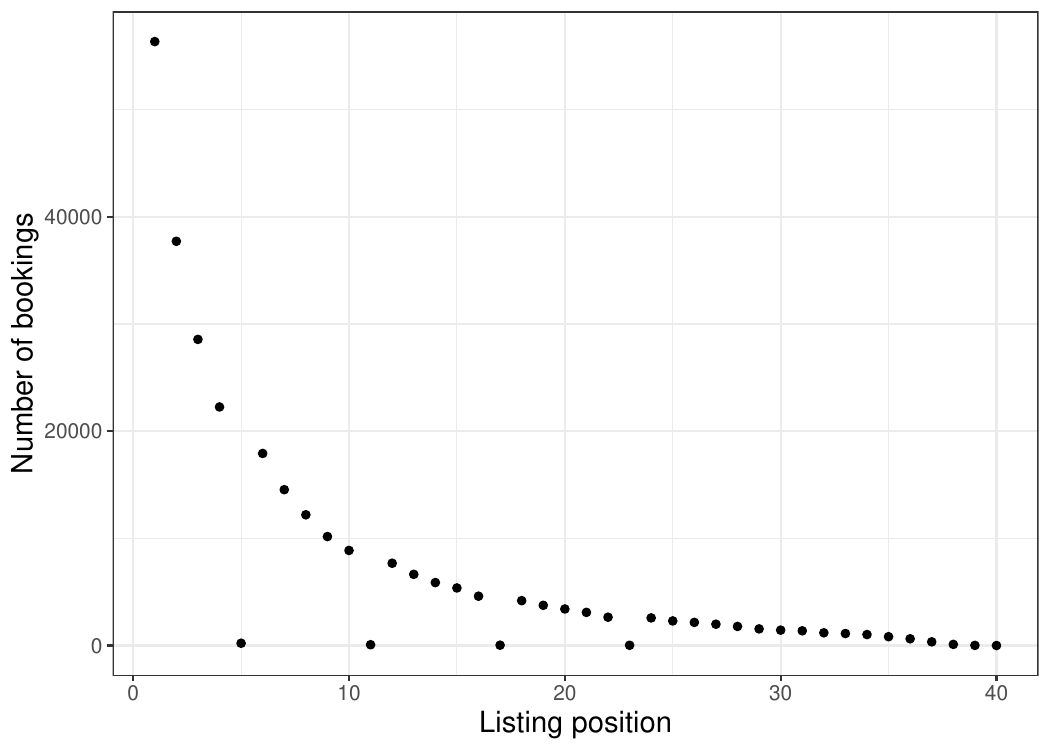}
    \caption{Total number of hotel bookings of each listing position in the ranked search results.}
    \label{fig:num_booking_per_position}
        \end{subfigure}  
        \begin{subfigure}[b]{0.4\textwidth}
            \centering
            \includegraphics[width=.8\textwidth]{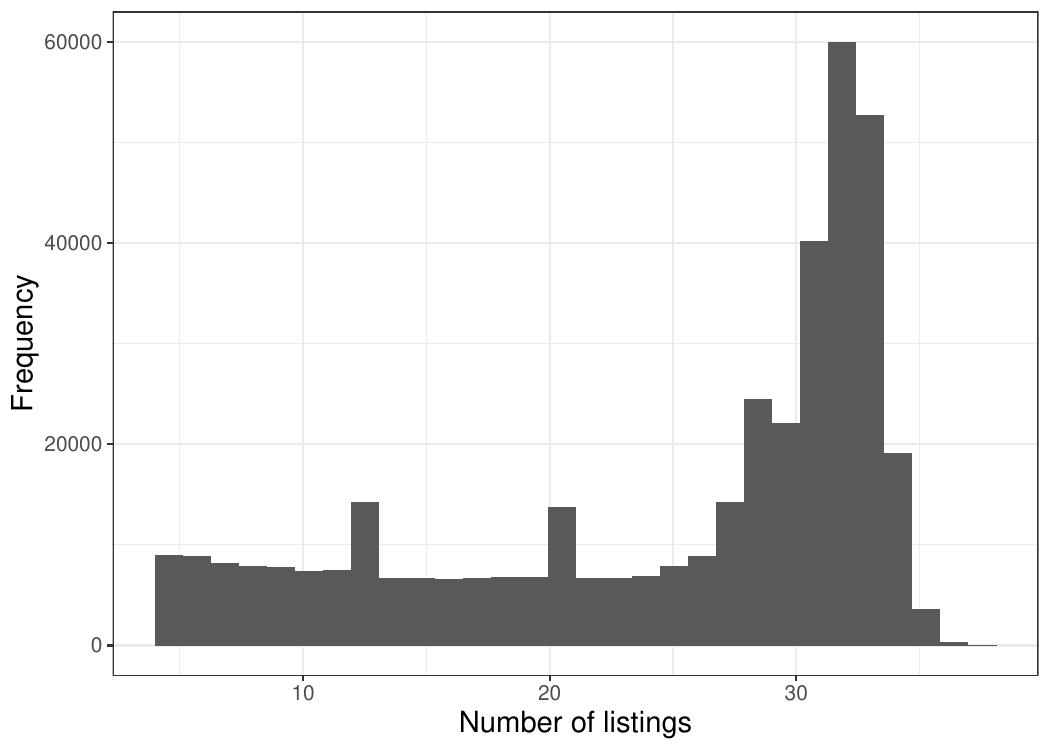}
    \caption{Distribution of numbers of search results per impression.}
     \label{fig:distr_num_listings}
        \end{subfigure}
        \caption{Statistics of hotel bookings. Note that several positions, e.g., 5, 10, have very few bookings. These are the ``mystery hotels'' which are not revealed to the customers upon booking, but determined later based on deals negotiated between Expedia and hotels.}
    \end{figure}

In the Expedia hotel search dataset, customers are provided a \emph{ranked} list, also known as an ``impression'', of search results based on their search inquiries for hotel bookings. The dataset contains hotel-level information, such as price, location score, star rating, as well as booking-specific information, such as the length of stay, number of guests, and how many days in the future is the searched booking. Importantly, in addition to impressions whose orders are determined by the platform's existing algorithm, the data also contain impressions where hotels are \emph{randomly} sorted. This randomization provides the foundation for identifying the positional effects in search results on Expedia's platform. Most impressions consist of between 5 to 35 listings, and in the data about 69\% of the searches resulted in a booking. 

A challenge with using the Expedia data is that each observation has a potentially different number of alternatives. However, as one might expect, most bookings involve top-ranked options, as customers' attention is limited and searching is costly. Figure~\ref{fig:num_booking_per_position} shows that very few bookings happen to options ranked below the 25th spot. Moreover, most of the impressions consist of at least 25 options, as Figure~\ref{fig:distr_num_listings} shows. We therefore process the dataset to include only impressions with at least 25 options, and \emph{truncate} the ordered lists by discarding options ranked lower than the 25th spot. This procedure results in observations with exactly 25 displayed alternatives. When a booking does not happen to any of the 25 alternatives, we treat this observation as having the ``default'' alternative as the outcome, which either means no booking was made, or a booking was made but with a listing ranked lower than 25. This procedure only reduced the percentage of searches that resulted in a booking from 69\% down to 67\%. Moreover, in principle we can apply a probit model with a larger response variable dimension to a sample with smaller choice set, by simply truncating the covariance matrix to the corresponding positions. 

Sometimes, customers are provided with very few ($<=5$) choices because of the restrictive search filters they applied or the low availability of hotels at particular times and locations. When the choice set is much smaller, we may expect that their decision making process is different from when the choice set has say 25 options. We may therefore also estimate another class of probit models, using only impressions with between 5 and 10 options, and truncating the impression to have exactly 5 options in the same way as before. If the researcher is interested in aggregating over impressions with different choice set sizes, the probit approach can in principle accommodate that as well. More precisely, if we specify a probit model with a response variable of dimension $m=25$, then for impressions with choice set size less than 25, we can truncate the probit model's covariance matrix, and then use the EM approach to update this subset of parameters only. 

\begin{figure}
    \centering
        \begin{subfigure}[b]{0.3\textwidth}
            \centering
            \includegraphics[width=\textwidth]{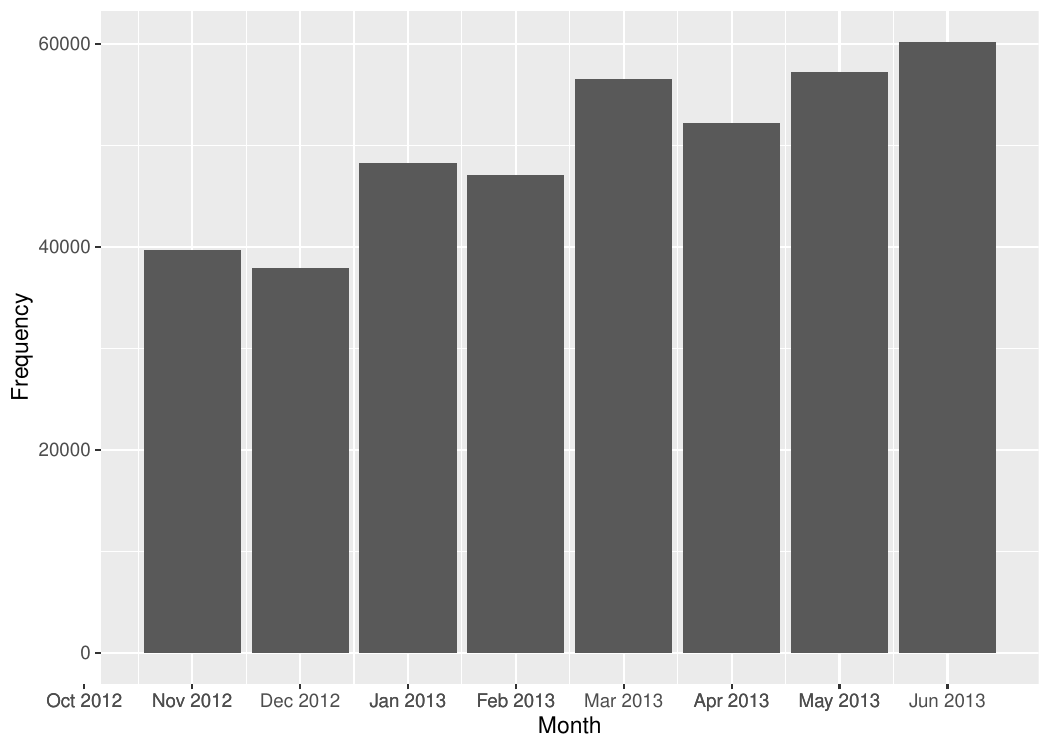}
            \caption{Distribution of numbers of searches per month.}
            \label{fig:search_month_distr}
        \end{subfigure}  
        \begin{subfigure}[b]{0.3\textwidth}
            \centering
            \includegraphics[width=\textwidth]{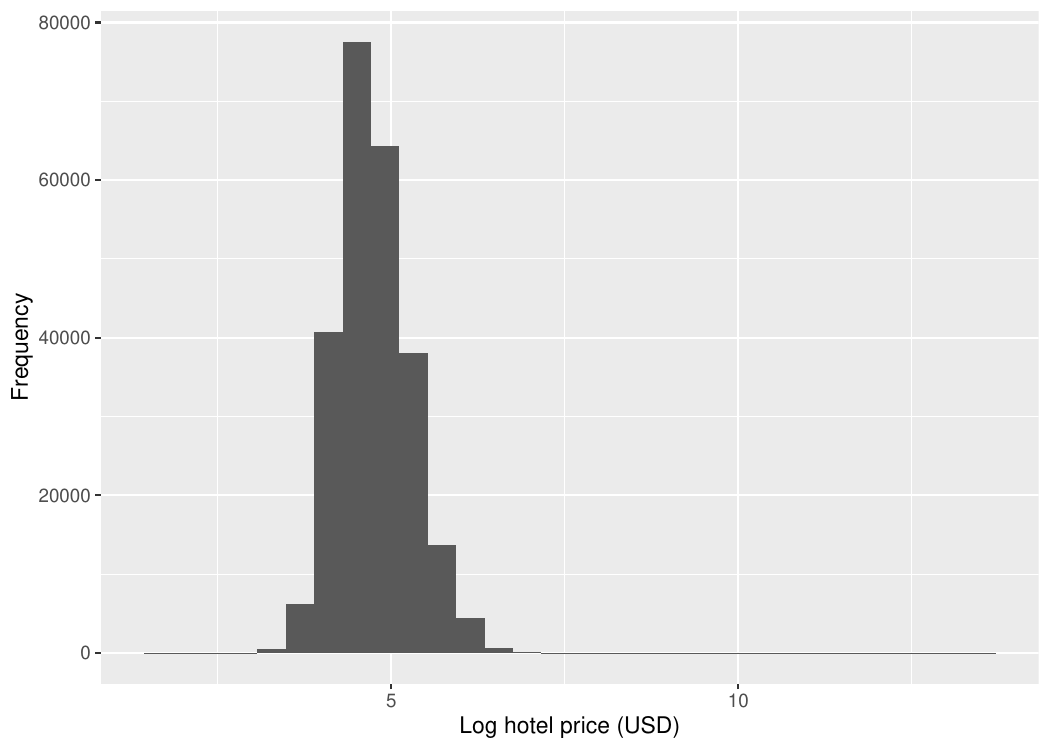}
            \caption{Log hotel price for US hotels in March 2013 for one night.}
            \label{fig:distr_log_price_us_march}
        \end{subfigure}
        \begin{subfigure}[b]{0.3\textwidth}
            \centering
            \includegraphics[width=\textwidth]{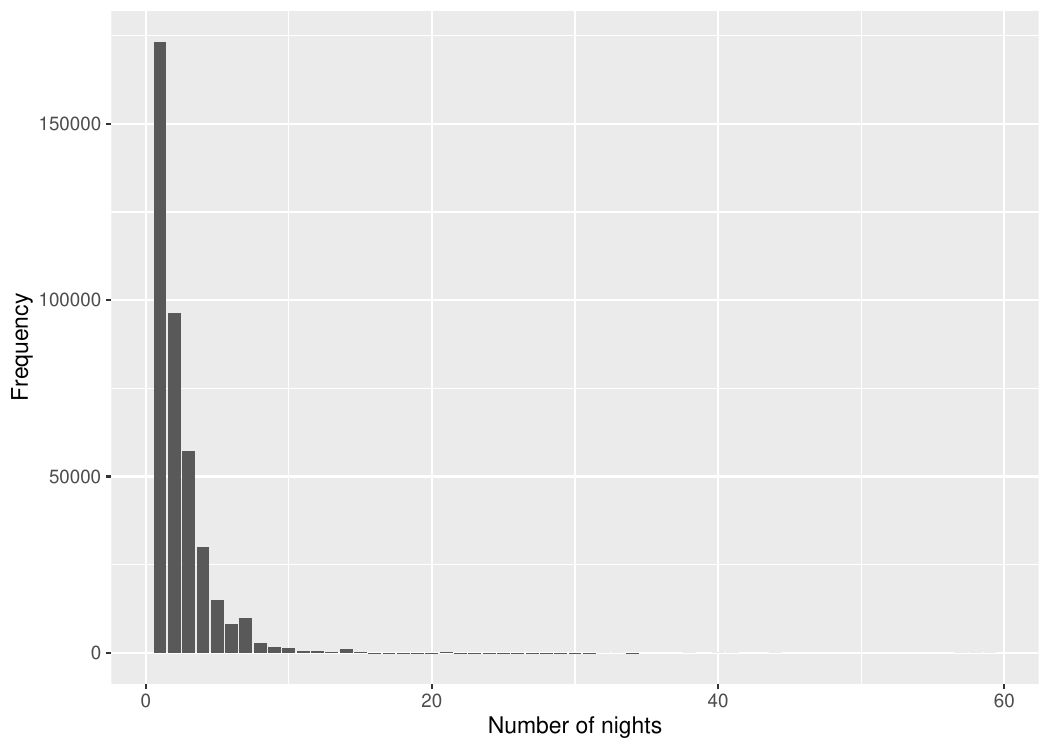}
            \caption{Number of searches of different stay lengths.}
            \label{fig:stay_length_distr}
        \end{subfigure}
        \caption{Statistics of hotel bookings. We restrict analysis to the most frequent types of searches, such as stays of 1 or 2 nights.}
    \end{figure}

The Expedia dataset contains property-specific information about the listings in each search result, such as price, the country it is in, average star ratings, and desirability scores of its location. In addition, it also contains search-specific information supplied by the customer, such as the number of nights, how many days in advance is the booking, and number of adults and children. When fitting the probit models, we use the following variables as covariates for each property: \texttt{prop\_starrating},  the average star rating, \texttt{prop\_brand\_bool}, indicator of whether the property is part of a major hotel chain, \texttt{prop\_location\_score1}, desirability of location, \texttt{price\_usd}, displayed price of the property, and \texttt{promotion\_flag}, indicator of whether the hotel has a sale price promotion displayed. We do not use search-specific information as covariates of the probit model, because we believe that the covariance structure could depend on the particular type of search, e.g., short stays vs. longer stays. Instead, we use such information to segment the data into subpopulations, on which we estimate separate probit models. 

The Expedia dataset has previously been used by \citet{abaluck2021consumers} in the context of consideration set models, where each item in a bundle has a certain probability of being considered by a consumer when they make decisions with limited attention or search costs. Their model can be used to decompose an item's demand into this attention effect as well as the usual utility effect. \citet{abaluck2021consumers} apply their model to predict the demand increase for items in the 3rd to 10th search positions when they are switched to the 1st or 2nd positions. Compared to their approach, we effectively model attention using correlated random utilities of a probit model. It would be interesting to compare these two approaches in more detail. 

\subsection{Results and Discussions}
\begin{figure}
    \centering
    \includegraphics[width=.6\textwidth]{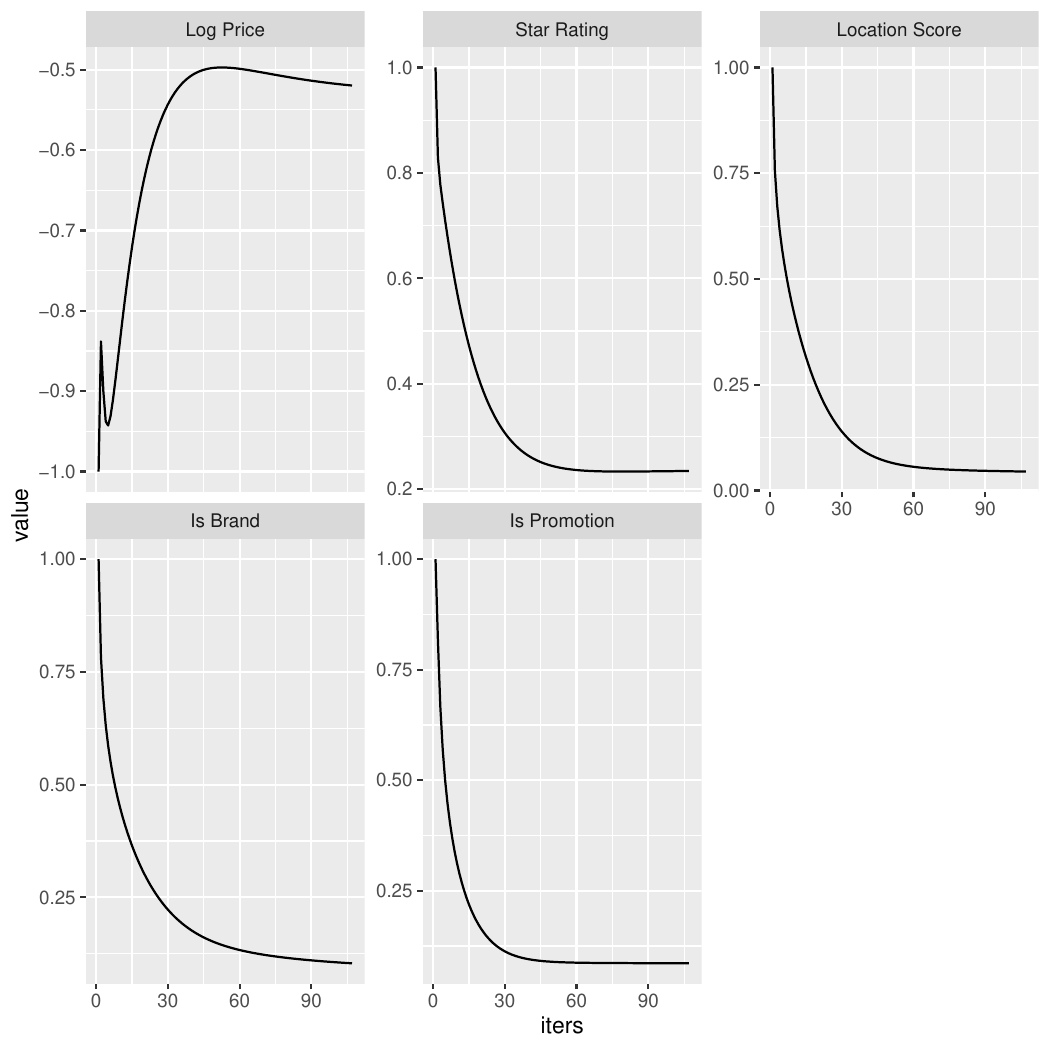}
    \caption{Estimated coefficients at each iteration in the probit model estimation process.}
    \label{fig:coeff-converge}
\end{figure}
We first present the estimated probit model based on searches in the third week of March 2013 for hotels located in the United States with a stay length of 1 night and 2 or less total guests. This sampling procedure results in 9856 impressions. We chose to focus on this subset of searches because the hotel price data for longer stays could contain total price or price per night, whereas for stays of 1 night the price is unambiguous. Moreover, we use the log price instead of the price because it has a more normal distribution (\cref{fig:distr_log_price_us_march}), and results in a more well-behaved probit model. In \cref{fig:coeff-converge}, we plot the histories of coefficient estimates over the EM algorithm iterations. Convergence is declared when the maximum absolute change in the covariance estimates is below a threshold. As expected, log price has a negative effect on the fixed component of the latent utility, whereas star rating, location score, brand recognition, and promotion all positively impact the fixed utility. 

\begin{figure}
    \centering
\includegraphics[width=.4\textwidth]{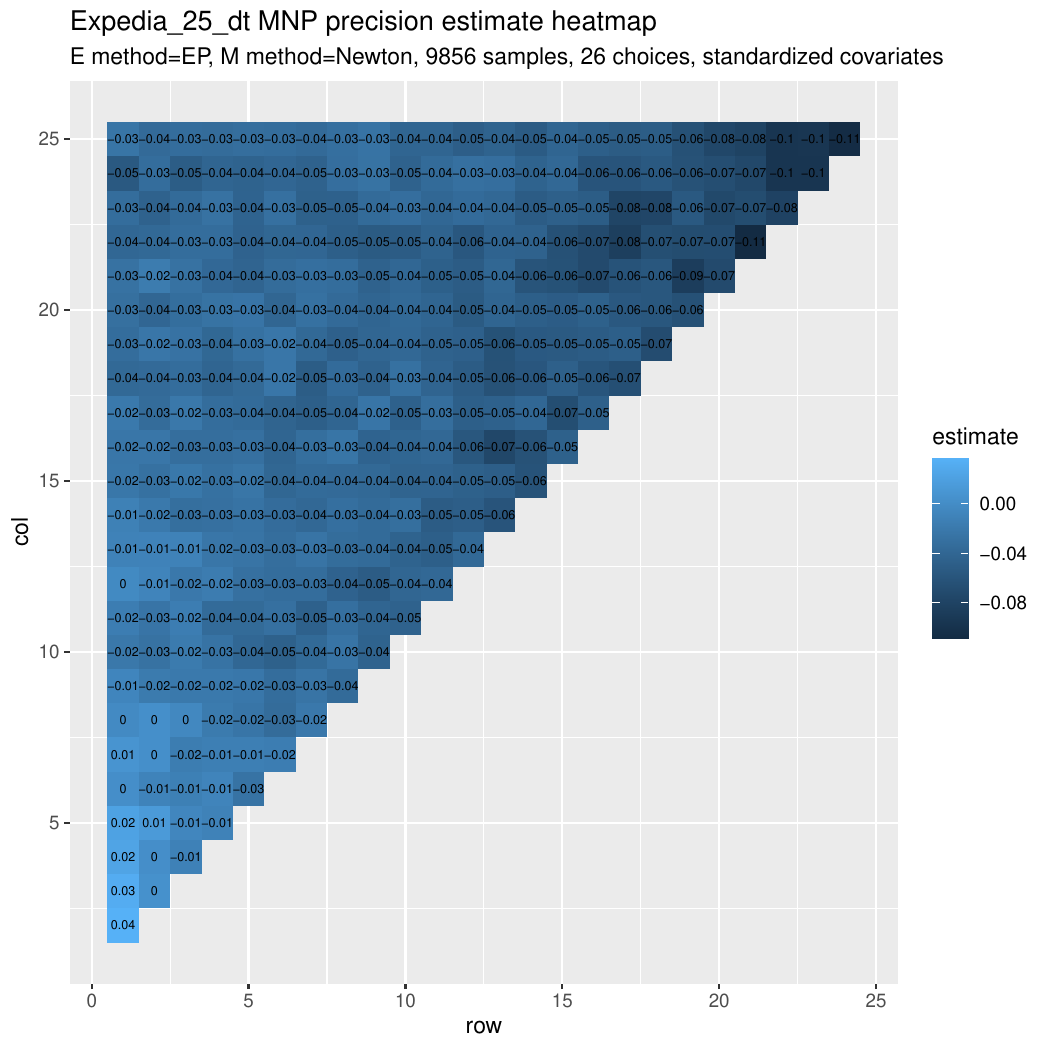}
    \caption{Heat map of the estimated precision matrix. The colored off-diagonal entries represent \emph{conditional} correlations of latent utilities, holding all other coordinates fixed. One observation is that the top few positions (lower left corner) behave as conditional complements to each other, while the bottom positions (upper right corner) behave as conditional substitutes.}
    \label{fig:precision-matrix-heat-map}
\end{figure}

\begin{figure}
    \centering
\includegraphics[width=.4\textwidth]{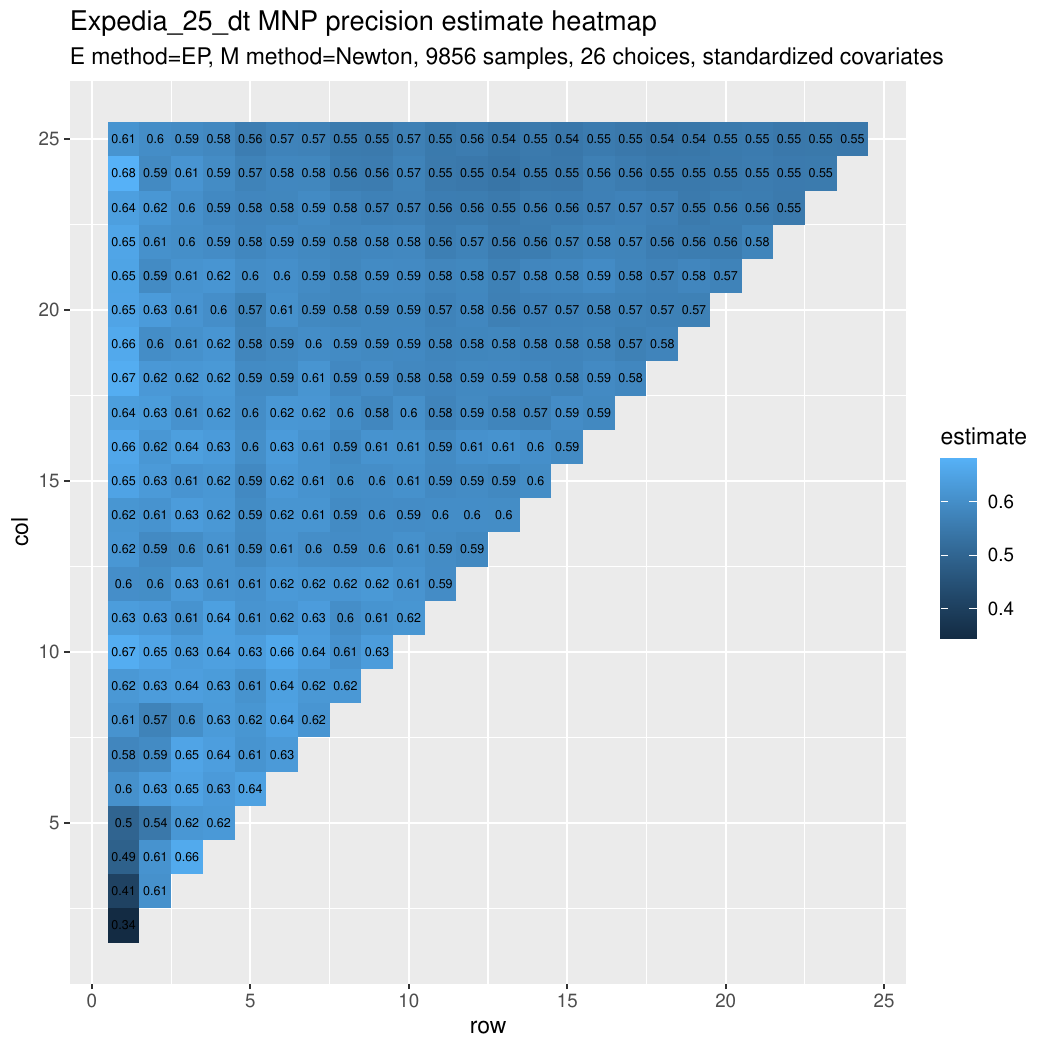}
    \caption{Heat map of the estimated covariance matrix. The colored off-diagonal entries represent \emph{marginal} correlations of latent utilities. The top few positions (lower left corner) have lower correlation, while the other positions have higher correlation.}
    \label{fig:covariance-matrix-heat-map}
\end{figure}

In \cref{fig:precision-matrix-heat-map}, we plot the heat map of the precision matrix of the estimated probit model. Recall that off-diagonal terms in the precision matrix represent \emph{conditional} correlations among coordinates of the latent multivariate normal utilities, which in this case correspond to different positions in an impression. We choose to examine the conditional correlations first, as we expect positional effects to be largely local: when customers make decisions, they tend to focus their attention on a small subset of adjacent options at a time. From \cref{fig:precision-matrix-heat-map}, we see that there are clear trends in the conditional correlations among different positions. For example, the top 5 positions have mostly positive conditional correlations, suggesting that their positional effects are complementary. On the other hand, the bottom 5 to 10 positions have negative conditional correlations. These observations are reasonably aligned with how customers make decisions on Expedia, and fit well in a limited attention framework. Initially, they spend a considerable amount of time and attention on the few top search results, and make careful comparisons among them. Thus the random utilities resulting from being placed in the top few spots in a ranked list should be positively correlated conditional on the realized random utilities of the other positions. As a consequence, if we switch the positions among the top few search results in the impression while fixing other items and their positions fixed in the bundle, we would expect smaller changes in their demands on average, since the customer would consider them carefully regardless. 
On the other hand, by the time a customer gets to the bottom of the list, they tend to become impatient and spend less effort investigating individual options. As a result, they make faster but less careful decisions, leading to cannibalization among the bottom options. We also plot the heat map of the estimated covariance matrix in \cref{fig:covariance-matrix-heat-map}. The marginal correlations between the top position and the 2nd to 4th positions are significantly smaller than the rest. This observation provides evidence that the top spot is indeed unique in terms of promoting the demand of search results, even compared to the other top spots in an impression. We verified that these observations are consistent across different subpopulations of the data, e.g., in different countries and months. 

\begin{figure}
    \centering
        \begin{subfigure}[b]{0.32\textwidth}
            \centering
            \includegraphics[width=\textwidth]{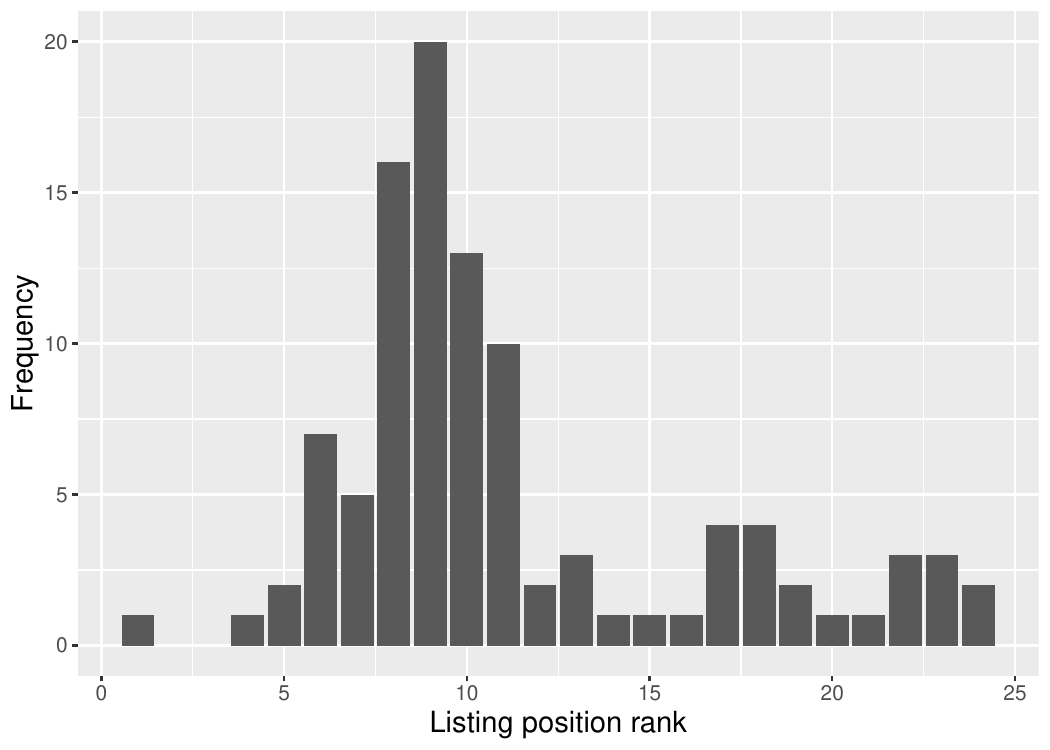}
            \caption{Distribution of positions of property 56880 in search results.}
        \end{subfigure}  
        \begin{subfigure}[b]{0.32\textwidth}
            \centering
            \includegraphics[width=\textwidth]{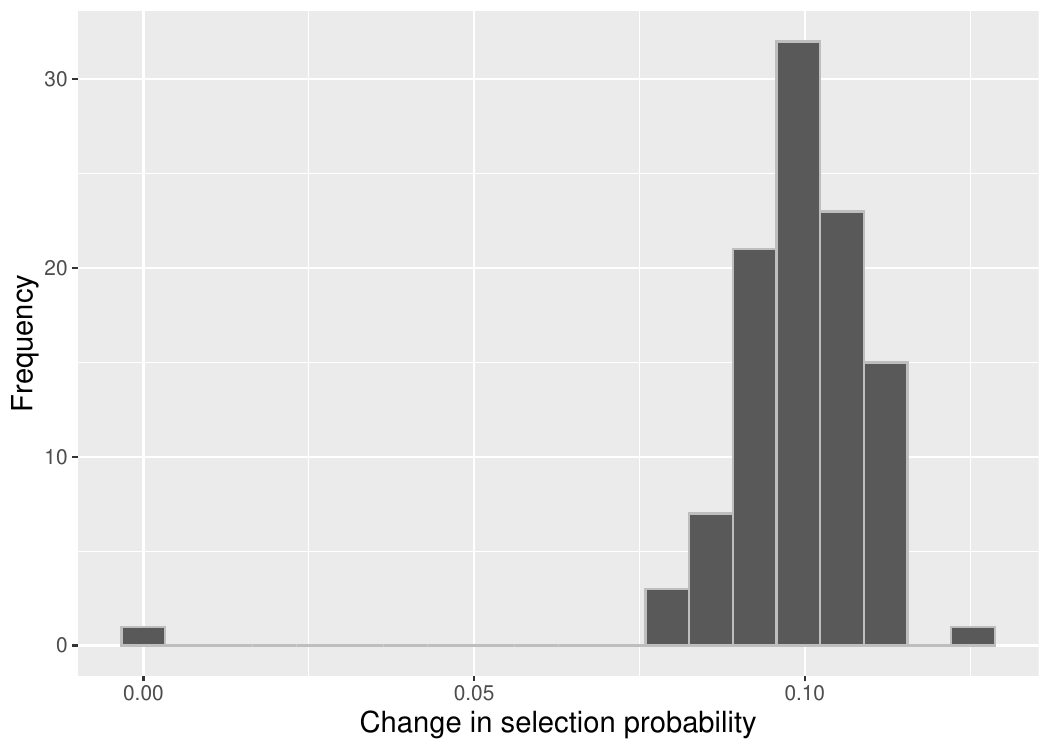}
            \caption{Distribution of predicted demand increases.}
        \end{subfigure}
        \begin{subfigure}[b]{0.32\textwidth}
            \centering
            \includegraphics[width=\textwidth]{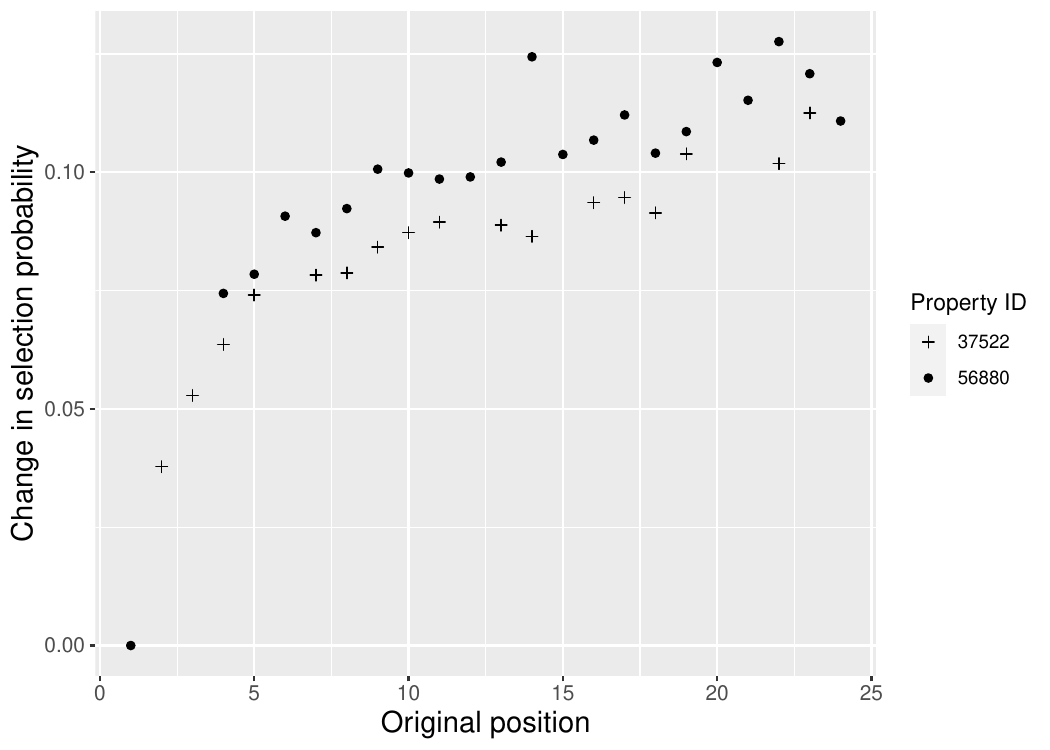}
            \caption{Predicted demand increase vs. original position.}
        \end{subfigure}
        \caption{Original positions of property 56880 in search results in March 2013 for one night stays of one or two adults (a), increases in booking probabilities by placing the property on top of the search list, predicted by the estimated probit model (b), and the correlation between the two (c). We observe a clear positive correlation when the original position of property 56880 in the search results is in the top 11 spots, and the correlation is weaker when the property is originally ranked lower.}
        \label{fig:counterfactual-56880}
    \end{figure}

\begin{figure}
    \centering
    \includegraphics[width=.4\textwidth]{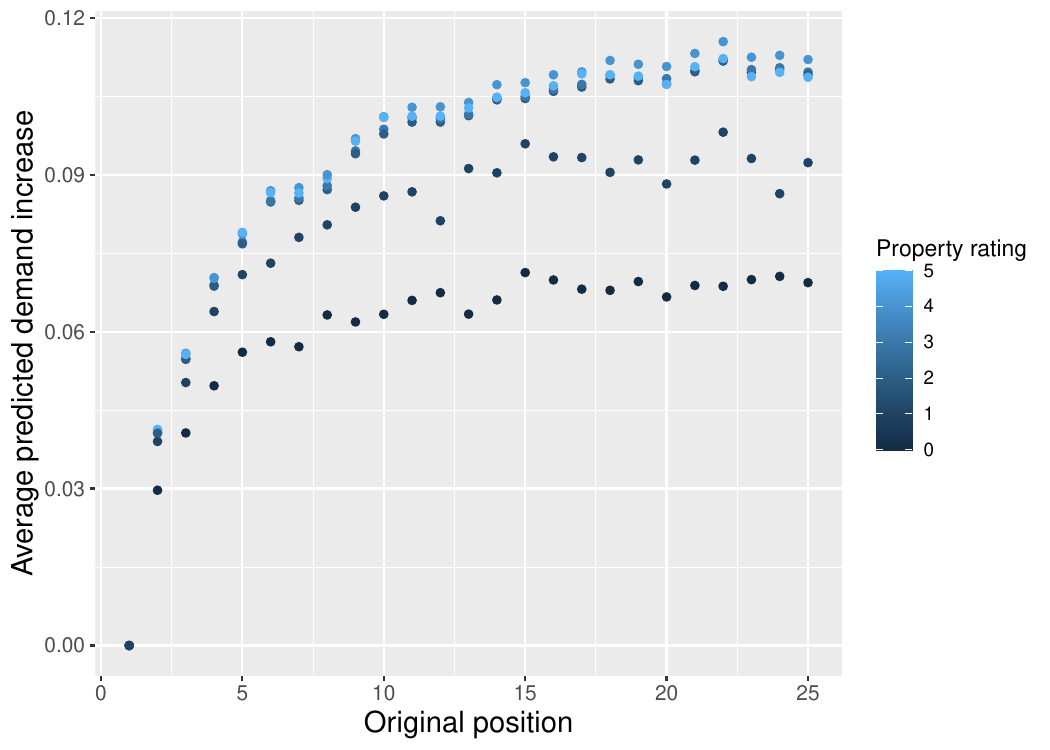}
    \caption{Estimated change in selection probability from placing a property in the top (sponsored) spot versus its original position. Color corresponds to rating of the property.}
    \label{fig:prob_change_starrating}
\end{figure}

Now we turn our attention to the motivating question laid out at the beginning of the section: how do we estimate the average increase in demand of a property when it is displayed on top of the search results instead of its usual positions? To answer this question, we can leverage the estimated probit model. More precisely, given a set of impressions containing the property of interest, we can first compute the choice probabilities of the property using the original ordering of alternatives in each impression. Then we modify the impressions by switching the position of the property to the top, which only involves switching coordinates of the fixed utility vector of the probit model. We can then compute the counterfactual choice probabilities using these modified impressions. Averaging the predicted demand increases over all impressions containing the property of interest then yields the predicted average demand increase. \cref{fig:counterfactual-56880} contains results for property 56880 in searches in March 2013 for stays of one night for one or two adults. We see that property 56880 most often appears in positions 5-10 in the search results we are interested in. If it were placed on top of the search results in these searches, our probit model predicts an average increase in demand of over 10\%. Moreover, the scatter plot of predicted probability increase vs. original position reveals that the lower the original rank of the property, the larger the increase in demand, although this trend is much more salient when the property is originally ranked in the top ten positions. We also found that, as expected, some properties benefit a lot more from sponsored search positions than others. For example, \cref{fig:prob_change_starrating} shows the average predicted demand increases for properties with different star ratings in March 2013. A salient observation is that for properties with original positions between 1 and 10 (likely on the first page of search results), highly-rated properties benefit even \emph{more} from increased exposures when placed in sponsored spots. This is somewhat counter-intuitive, as highly rated properties are more likely to be placed closer to the top of the ranked list, so we may expect the effects of being placed in sponsored spots to be diminishing. An immediate implication is that highly rated properties can still benefit significantly from sponsored displays, while some lower rated properties are unlikely to benefit from sponsored displays without improving their qualities. More broadly, the heterogeneous benefits of sponsored spots for different properties is an interesting feature of the data that our probit approach is able to capture compared to other standard approaches. These results demonstrate the practical usefulness of our probit estimation framework for datasets with ordered choice sets. Moreover, our approach is applicable to a wide range of applications, such as streaming platforms, online shopping platforms, and geospatial problems, and is scalable to choice set sizes over 200. 







\section{Conclusion}
\label{sec:conclusion}
In this work, we propose a computationally efficient EM algorithm for the estimation of large probit models. Our main innovations are twofold. First, we propose to circumvent the computational challenges of simulation and sampling based methods by using expectation propagation to approximate the moments of TMVN in the E step. Second, we propose a symmetric trace identification constraint which enables us to reduce the constrained optimization problem in the M step to that of finding the root of a one-dimensional monotonic function. This result highlights interesting interactions between the statistical issue of identification and the optimization problem of maximum likelihood. We demonstrate that the resulting EM algorithm is comparable or outperforms in accuracy with existing methods on small problems, but is able to estimate much larger probit models. We demonstrate the practical appeal of our method on the Expedia hotel search data, where choice sets are ranked. We find that the benefits of sponsored displays exhibits considerable \emph{heterogeneity} in both product characteristics and counterfactual positions. An immediate implication of our results is that highly rated properties can still benefit significantly from sponsored spots. These results demonstrate the practical usefulness of our probit estimation framework to capture contextual effects in datasets with ordered choice sets.

Our probit approach to estimate positional effects in choice data also applies to other settings with ordered choice sets. For example, online streaming services often organize offerings in a panel, and the position of a movie can affect the probability that a customer chooses to watch it. An \texttt{R} package that implements the proposed EM algorithm is under construction. For future works, it would be interesting to further study the convergence properties of our EP-based algorithm. Moreover, interactions between novel structural assumptions and the estimation of probit models provide exciting new research directions. 

 \section*{Acknowledgements}
 {We thank
the Office of Naval Research for support under grant numbers N00014-17-1-2131 and N00014-19-1-2468 and Amazon for a gift. This work was also supported in part by a Stanford Interdisciplinary Graduate Fellowship (SIGF). We are very grateful for helpful discussions with and comments and suggestions from Gonzalo Arrieta, Adrien Auclert, Doug Bernheim, Varanya Chaubey, Liran Einav, Alfred
Galichon, Peter Glynn, Han Hong, Matt Jackson, Boyan Jovanovic, Yongchan Kwon, Debdeep Pati, Brad Ross, Johan
Ugander, Quang Vuong, Ruxian Wang, participants of the
3rd Year Seminar, the EconScribe Writing Course, the Econometrics Lunch Seminar at Stanford Econ, and participants of the 2023 Stanford Data Science Conference.}

\bibliography{ref,Covariance_Estimation}
\bibliographystyle{apalike}

\newpage

\begin{appendices}
\section{Brief Review of Probit and Logit Models}
\label{sec:probit-logit-introduction}
Both probit and logit models were initially developed for one-dimensional
binary responses. 
More precisely, we observe the response variable
$Y\in\{0,1\}$ and covariates $X$. 
In this setting, the models can be formulated as generalized linear models (GLM), with likelihood given by 
\begin{align*}
\mathbb{P}(Y=1\mid X) & =\Phi(X^{T}\beta).
\end{align*}
Probit and logit differ in the choice of $\Phi$. 
In probit models $\Phi$ is the CDF of the standard normal distribution, while in logit models $\Phi$ is defined uniquely through the log-odds
\begin{align*}
\Phi^{-1}(\mathbb{P}(Y=1\mid X))=\log(\frac{\mathbb{P}(Y=1\mid X)}{1-\mathbb{P}(Y=1\mid X)}) & =X^{T}\beta,
\end{align*}
which requires using the logistic function 
\begin{align*}
\Phi(\eta) & =\frac{\exp(\eta)}{1+\exp(\eta)}.
\end{align*}

Probit and logit models can be understood as latent variable models.
Let 
\begin{align*}
Z & =X^{T}\beta+\epsilon\\
Y & =1\{Z\geq0\}.
\end{align*}
 Then the likelihood function 
\begin{align*}
\mathbb{P}(Y=1\mid X) & =\mathbb{P}(-X^{T}\beta\leq\epsilon)
\end{align*}
is that of the probit model when $\epsilon$ is distributed as a standard normal, and is that of the logit model when $\epsilon$ has Gumbel distribution.

Both probit and logit are readily extended to model multi-dimensional discrete response variables in $\{0,1\}^m$. For the probit model, we use a multi-dimensional normal latent variable 
\begin{align*}
    Z_{i} & \sim\mathcal{N}(X_{i}\beta,\Sigma),
\end{align*}
with unknown coefficient vector $\beta\in\mathbb{R}^{p}$ and covariance matrix $\Sigma$. There are different specifications of the relation between the observed response variable $Y_{i}$ and the unobserved latent variable $Z_{i}$. For example, panel data often assumes
\begin{align*}
    Y_{ij} & =
    \begin{cases}
        1 & \text{if }Z_{ij}\geq0\\
        0 & \text{otherwise}
    \end{cases}, 
\end{align*}
while discrete choice data assumes 
\begin{align*}
Y_{ij} & =\begin{cases}
1 & \text{if }Z_{ij}=\max_{j}Z_{ij}\\
0 & \text{otherwise}
\end{cases}.
\end{align*}

 For the discrete choice probit model, if we stack up the rows of $X_{i}$ into a column vector,
the latent variable model can be equivalently written as 
\begin{align*}
Z_{ij} & =X_{i}^{T}\beta_{j}+\epsilon_{ij},
\end{align*}
 where $\epsilon_{i}\sim\mathcal{N}(0,\Sigma)$, $\beta_{j}=e_{j}\otimes\beta$
where $e_{j}\in\mathbb{R}^{m}$ is the indicator vector with $1$
at the $j$th component, and $\otimes$ is the Kronecker product.
Conversely, the specification $Z_{ij}=X_{i}^{T}\beta_{j}+\epsilon_{ij}$ with alternative-specific $\beta_j$ can be rewritten as $Z_{ij}=X_{ij}^{T}\beta+\epsilon_{ij}$. These
two models allow different interpretations: the model with fixed
$\beta$ is used to model situations where $X_{ij}$, the $j$-th
row of $X_{i}$, corresponds to the characteristics of the $j$th
alternative in the $i$-th observation, and $\beta$ represents the
fixed effects of these characteristics. On the other hand, in the variable coefficient
formulation with $\beta_{j}$ corresponding to the coefficients of
alternative $j$, $X_{i}$ can include both the characteristics of
the decision maker, as well as the characteristics of the choices.
Another extension of the probit model assumes random coefficients
$\beta_{i}$ with normal distributions, in order to model the heterogeneity
in consumers' tastes. In this paper, we use the fixed coefficient
model
\begin{align*}
Z_{ij} & =X_{ij}^{T}\beta+\epsilon_{ij}
\end{align*}
because our focus is on the covariance matrix $\Sigma$, although
the approaches we develop in this paper apply to both types of specifications, and can be generalized to the random coefficient model as well.

The multinomial probit model is invariant under an observation-specific shift of $Z_{ij}$. The following procedure \citep{train2009discrete} is routinely used to ensure identification of the probit model. We can subtract $Z_{i1}$, the latent utility
of the first alternative, from $Z_{ij}$ for all $j$, to obtain an
$m-1$ dimensional \emph{random} vector $\{\tilde{Z}_{ij}\}_{j=2}^{m}$: 
\begin{align*}
\tilde{Z}_{ij} & =Z_{ij}-Z_{i1}\\
 & =(X_{ij}-X_{i1})^{T}\beta,
\end{align*}
The transformed latent variables now satisfy 
\begin{align*}
Y_{ij} & =\begin{cases}
1 & \text{if }\tilde{Z}_{ij}=\max_{j>1}\tilde{Z}_{ij}\text{ and }\tilde{Z}_{ij}>0\\
0 & \text{otherwise}
\end{cases}
\end{align*}
for $j>1$ and 
\begin{align*}
Y_{ij} & =\begin{cases}
1 & \text{if }\tilde{Z}_{ij}\leq 0\text{ for all }j>1\\
0 & \text{otherwise}
\end{cases}
\end{align*}
 for $j=1$. We can readily obtain the covariance matrix $\tilde{\Sigma}$
 of this $m-1$ dimensional system from $\Sigma$. The reduced model parameterized by $(\beta,\tilde{\Sigma})$ 
with normalized covariates $\tilde{X}_{ij}=X_{ij}-X_{i1}$ for $j>1$
generates the same distribution on $Y_{i}$ as the original model
parameterized by $(\beta,\Sigma)$. Therefore, we may instead identify the reduced system with $(\beta,\tilde{\Sigma})$. 

\section{Introduction to Expectation Propagation (EP)}
\label{sec:EP-detailed-intro}
In this section we provide a self-contained introduction to EP which is a key component in our EM algorithm for probit estimation.
\subsection{Intuition of EP}
The method proposed in \citet{cunningham_gaussian_2013} applies an approximate Bayesian inference algorithm to compute truncated multivariate normal density integrals. In Bayesian inference, we are interested in probabilistic queries involving the posterior density over latent variables $z$ given data $x$,
\begin{align}\label{eq:posterior}
    p(z|x) = \frac{p(x|z)p(z)}{p(x)}.
\end{align}
For example we may want to compute posterior expectations $\int z p(z|x)dz$.
The challenge is that \eqref{eq:posterior} is often intractable because of the marginal likelihood $p(x)$. 
Markov chain Monte Carlo (MCMC) methods avoid this problem by drawing samples from $p(z|x)$ without needing knowledge of $p(x)$, which can be used for Monte Carlo estimates of posterior quantities of interest.
MCMC methods will eventually sample from the true posterior after enough iterations.
However, they are computationally intensive.

\begin{figure}[h!]
    \centering
    \includegraphics[width=.4\textwidth]{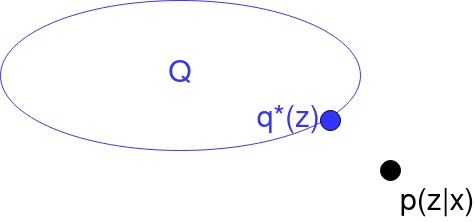}
    \caption{Variational inference; find $q(z) \in \mathcal{Q}$ that is in some sense closest to $p(z|x)$.}
    \label{fig:vi_idea}
\end{figure}

Variational inference methods \citep{bleiVariationalInferenceReview2017} are a more efficient, albeit approximate, alternative to MCMC methods. 
The idea is to find a density $q^*(z)$ from a tractable family $\mathcal{Q}$ that best approximates $p(z|x)$ (Figure~\ref{fig:vi_idea}).
$q^*(z)$ is determined by minimizing a divergence $D$ between the true posterior $p(z|x)$ and approximation $q(z) \in \mathcal{Q}$, where $\mathcal{Q}$ is a family of distributions over $z$:
\begin{align}\label{eq:vi_objective}
    q^*(z) = \min_{q \in \mathcal{Q}} D(p(z|x) || q(z) ).
\end{align}

We will need to specify the approximating family $\mathcal{Q}$ and the choice of divergence $D$.
The approximating family $\mathcal{Q}$ needs to trade off flexibility, which allows accurate approximation, with computational speed. 
The choice of divergence $D$ also has an impact on the approximation. 
Variational bayes uses the reverse KL divergence,
\begin{align}\label{eq:reverse_kl}
    D_{KL}(q(z) || p(z|x)) = \int q(z) \log \frac{ q(z) }{p(z|x)} dz
\end{align}
which encourages mode seeking approximations. 
The forward Kullback-Leibler (KL) divergence
\begin{align}\label{eq:forward_kl}
    D_{KL}(p(z|x) || q(z)) = \int p(z|x) \log \frac{ p(z|x) }{q(z)} dz
\end{align}
is an inclusive divergence that promotes approximations $q^*$ that cover as much of the nonzero probability regions of $p(z|x)$ as possible, and also well approximates the normalizing constant of $p(z|x)$ \citep{minkaDivergenceMeasuresMessage2005}.

However, this objective is still intractable since it involves $p(z|x)$. 
Expectation propagation (EP) \citep{minka2001expectation,minka2001family} is a method that circumvents this issue by minimizing \emph{local} forward KL divergences.
It sets $\mathcal{Q}$ to be the family of multivariate normal distributions. 
This combination of choices leads to an approximation that captures the zeroth (normalizing constant), first, and second moments of the target posterior, which is exactly what we need for the truncated multivariate normal distribution in probit estimation. \citet{cunningham_gaussian_2013} use EP to approximate truncated multivariate normals with unconstrained multivariate normal distributions (Figure~\ref{fig:ep_orthant_tmvn}), but the application of EP in the estimation of probit models is uncommon, appearing only in \citet{mandt_sparse_2017} and in \citet{fasano2023efficient} in a Bayesian framework.

\begin{figure}[h!]
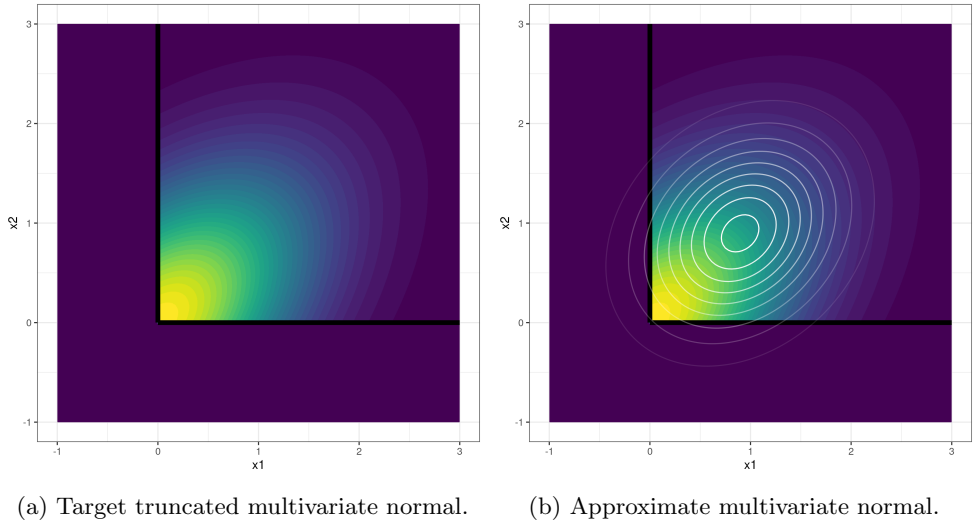

    \centering
    \begin{subfigure}[b]{0.4\textwidth}
        \centering
        \includegraphics[width=\textwidth]{ep_exposition/target_tmvn.pdf}
        \caption{Target truncated multivariate normal.}
    \end{subfigure}
    \begin{subfigure}[b]{0.4\textwidth}
        \centering
        \includegraphics[width=\textwidth]{ep_exposition/approx_mvn.pdf}
        \caption{Approximate multivariate normal.}
    \end{subfigure}
    \caption{EP multivariate normal approximation to positive orthant truncated multivariate normal.}
\end{figure}

\subsection{The EP Method}

\citet{cunningham_gaussian_2013} derives a deterministic expectation propagation approximation for estimating truncated multivariate normal (TMVN) first and second moments under linear domain constraints. 
The idea is to approximate the target TMVN with an un-truncated multivariate normal (MVN) that has tractable moments.

Let the target $d$-dimensional TMVN density of interest be $p(x)$, where
\begin{align}\label{eq:ep_target_mvn}
    p(x) = \frac{1}{\zeta} \exp \left(-\frac{1}{2}(x - \mu)^T \Sigma^{-1}(x-\mu)  \right) \mathbbm{1}_{l < Ax < u}(x),
\end{align}
and $\mu$ is the mean, $\Sigma$ is the covariance, $\zeta$ is the normalizing constant, $l,u \in \mathbb{R}^{d}$ are the lower and upper bounds, and $A \in \mathbb{R}^{d\times d}$ is the linear constraint matrix.
We rewrite $p(x)$ as a product of factors:
\begin{align}\label{eq:ep_target_factorization}
    p(x) &= p_{0}(x) \prod_{k=1}^{d} t_{k}(x)
\end{align}
where $p_0(x) = \frac{1}{\zeta} \exp \left(-\frac{1}{2}(x - \mu)^T \Sigma^{-1}(x-\mu)  \right)$, and 
\begin{align}\label{eq:ep_target_factor}
    t_k(x) = \mathbbm{1}_{l_k < a_k^Tx < u_k}(x),
\end{align}
where $a_k$ is the $k$th row of $A$ that parameterizes the $k$th linear constraint.

We want to approximate $p(x)$ with $q(x)$, where $q(x)$ has the same factorized form as $p(x)$:
\begin{align}\label{eq:ep_approx}
    q(x) = p_0(x) \prod_{k=1}^d \tilde{t}_k(a_k^T x),
\end{align}
and 
\begin{align}\label{eq:approx_factor}
    \tilde{t}_k(x) = \mathcal{N}(a_k^T x, \tilde{\mu}_k, \tilde{\sigma}_k^2).
\end{align}
$p_0(x)$ and the $\tilde{t}_k(x)$ are all normal densities, so the product of their densities will also be normal. 
Therefore $q(x)$ has a multivariate normal density,
\begin{align}\label{eq:approx_mvn}
    q(x) = \mathcal{N}(x; \tilde{\mu}, \tilde{\Sigma}).
\end{align}
Our goal is to find $\tilde{\mu}$ and $\tilde{\Sigma}$ such that $q(x)$ is a good approximation to $p(x)$. 
Cunningham et al. use the expectation propagation algorithm \citep{minka2001expectation} to achieve this task for truncated normal approximation. 
We summarize the expectation propagation algorithm exposition of Cunningham et al. 
The algorithm iterates over the factors $k = 1, \ldots, d$:
\begin{enumerate}
    \item 
        Form \textit{cavity distribution} by removing the $k$th approximate factor from $q(x)$:
        \begin{align*}
    q^{\backslash k}(x) = \frac{q(x)}{\tilde{t}_k(x)}
        \end{align*}
    \item 
        Form the \textit{tilted distribution} by multiplying cavity distribution with the $k$th true factor:
        \begin{align*}
             t_{k}(x) q^{\backslash k}(x)
        \end{align*}
    \item 
        Update the $k$th approximate factor $\tilde{t}_k(x)$ such that it minimizes the KL divergence between the true target distribution and the tilted distribution:
        \begin{align*}
        \tilde{t}_k(x)^{new} = \arg \min D_{KL}(t_{k}(x) q^{\backslash k}(x) ||  \tilde{t}_{k}(x) q^{\backslash k}(x))
        \end{align*}
        Since $\tilde{t}_k(x)$ is normal (it is an exponential family distribution), this minimization is equivalent to matching the moments of $t_{k}(x) q^{\backslash k}(x)$ and $\tilde{t}_{k}(x) q^{\backslash k}(x)$.
    \item 
        Update the overall approximation, set $q(x) = \tilde{t}_k(x)^{new}q^{\backslash k}(x)$. This is a product of normal densities, so $q(x)$ is still normal and there are closed forms for its normalizing constant, mean, and covariance.
\end{enumerate}
After iterating through all the factors this procedures repeats, until $\tilde{\mu}$ and $\tilde{\Sigma}$ converge.
Upon termination, we use $\tilde{\mu}$ and $\tilde{\Sigma}$ as approximations of the mean and covariance of the TMVN $p(x)$.

Although EP does not have convergence guarantees, 
Cunningham et al observe that empirically it is fast and accurate for multivariate normal probability estimation, especially for rectangular constraints where $A = I_d$. 
They found some failure cases for EP, for example when there are many redundant constraints. Also, note that the forward KL divergence objectives in EP ($D_{KL}(p(x)|| q(x)$) are \emph{inclusive}, meaning the approximating distribution tries to cover all nonzero probability regions of the true target density. 
This is in contrast to the exclusive reverse KL divergence ($D_{KL}(q(x)|| p(x)$) minimization, meaning the approximating distribution seeks a mode of the target density. 
See \citet{minkaDivergenceMeasuresMessage2005}. The EP algorithm minimizes local forward KL divergences, instead of a global objective $D_{KL}(p(x)|| q(x))$. 
This is because minimization of the global objective is intractable.

\section{Accelerating EP for Probit Estimation}\label{sec:faster_ep}
The general EP algorithm applied to estimate moments of TMVN has complexity $\mathcal{O}(m^3)$. However, if the linear constraints that define the TMVN are axis-aligned, i.e., the constraint region is rectangular, then EP can be accelerated to have only $\mathcal{O}(m^2)$ dependence. It turns out that the structure of TMVNs resulting from probit models allows us to transform the problem in the E step to this simpler setting. More precisely, the latent utility $Z_i$ restricted to the linearly constrained region $l\leq A_i Z_i \leq u$ can be transformed into a variable $U_i$ restricted to an axis aligned region, via the transformation $U_i = A_i (Z_i - X_i\beta)$. Since the original constraints are $l\leq A_i Z_i \leq u$,
$U_i$ is also a TMVN subject to the rectangular constraints 
\begin{align*}
    l - A_i(X_i\beta) \le U_i \le u - A_i (X_i\beta),
\end{align*}
and has mean 0 and covariance $A_i \Sigma A_i^T$. Importantly, the involutory property of $A_i$ guarantees that we can easily transform back from $U_i$ to $Z_i$: 
\begin{align*}
   A_i U_i =   A_i A_i (Z_i-X_i\beta)=Z_i-X_i\beta,
\end{align*}
so that $Z_i =A_i U_i+X_i\beta$. This representation is crucial for the efficiency of the EP approximation, since its implementation for the axis-aligned problem with $U_i$ has quadratic complexity in choice set size $m$, compared to cubic complexity for the general problem with $Z_i$ \citep{cunningham_gaussian_2013}. Let $\mu_i'$ and $S'_i$ be the EP approximations to the mean and covariance in the transformed space, i.e., of the (conditional) moments of $U_i$. To recover the the moments $\mu_i,S_i$ of the TMVN $Z_i$ in the original space, we can simply compute
\begin{align}
    \mu_i = A\mu_i' + X_i\beta
    \\
    S_i = A_iS'_iA_i^T,
\end{align}
where we use the involutory property of $A_i$ to invert the transformation $U_i = A_i (Z_i - X_i\beta)$. The constraint matrix can be similarly constructed for the multivariate probit model. Our proposal is to use EP to approximate the TMVN moments in \cref{alg:em_generic}. Furthermore, since the moments can be computed separately for each sample, EP computations can be parallelized across observations. 

    \section{Additional Simulation Results for Different Specifications of the Covariance Matrix}
In this section, we present additional simulation results based on data generated from covariance matrix structures other than the compound symmetric structure used in \cref{sec:simulations}.
    
    \begin{figure}[h!]
        \centering
        \begin{subfigure}[b]{0.32\textwidth}
            \centering
            \includegraphics[width=\textwidth]{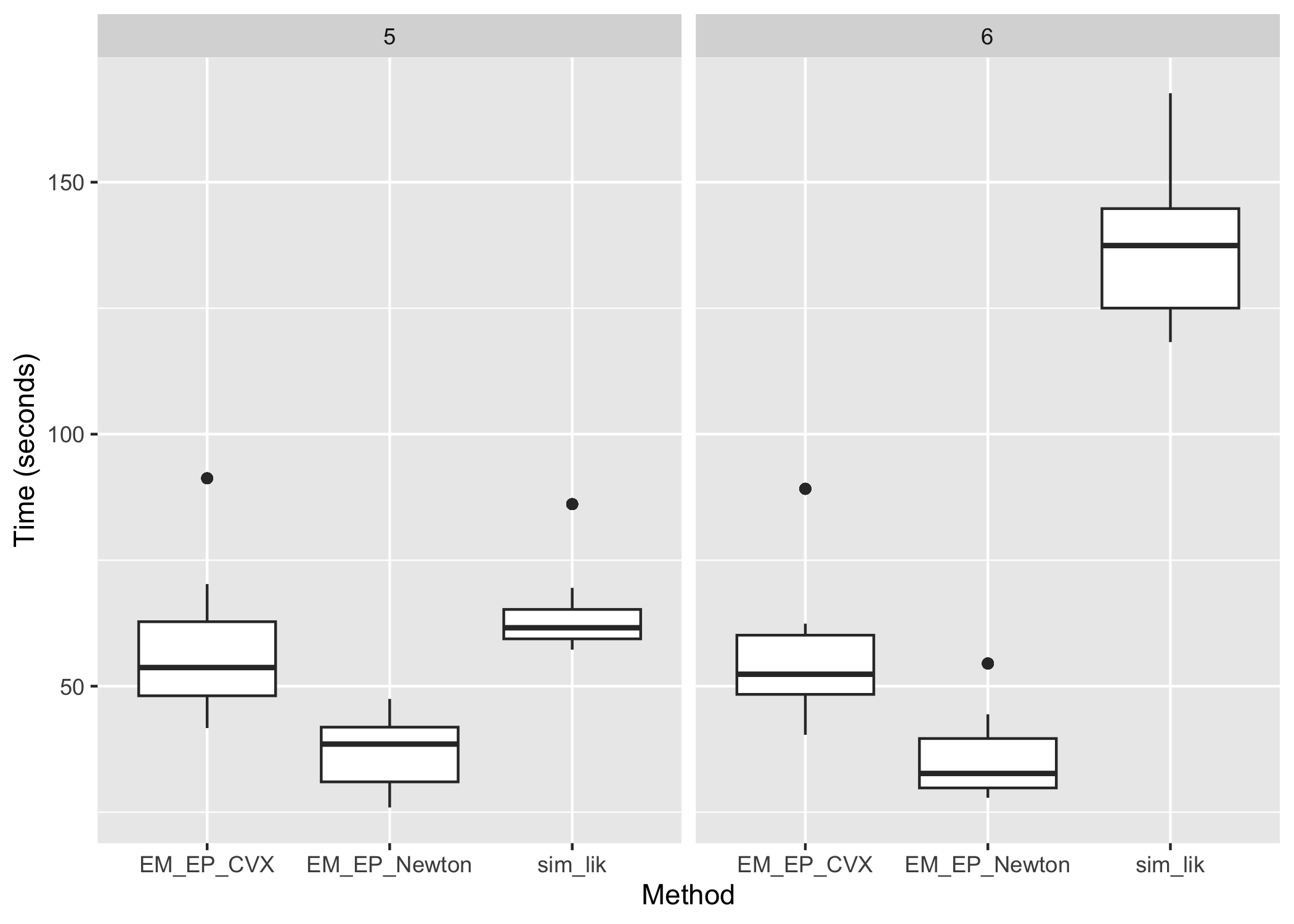}
            \caption{Runtime}
        \end{subfigure}
        \hspace*{.1\textwidth}
        \begin{subfigure}[b]{0.32\textwidth}
            \centering
            \includegraphics[width=\textwidth]{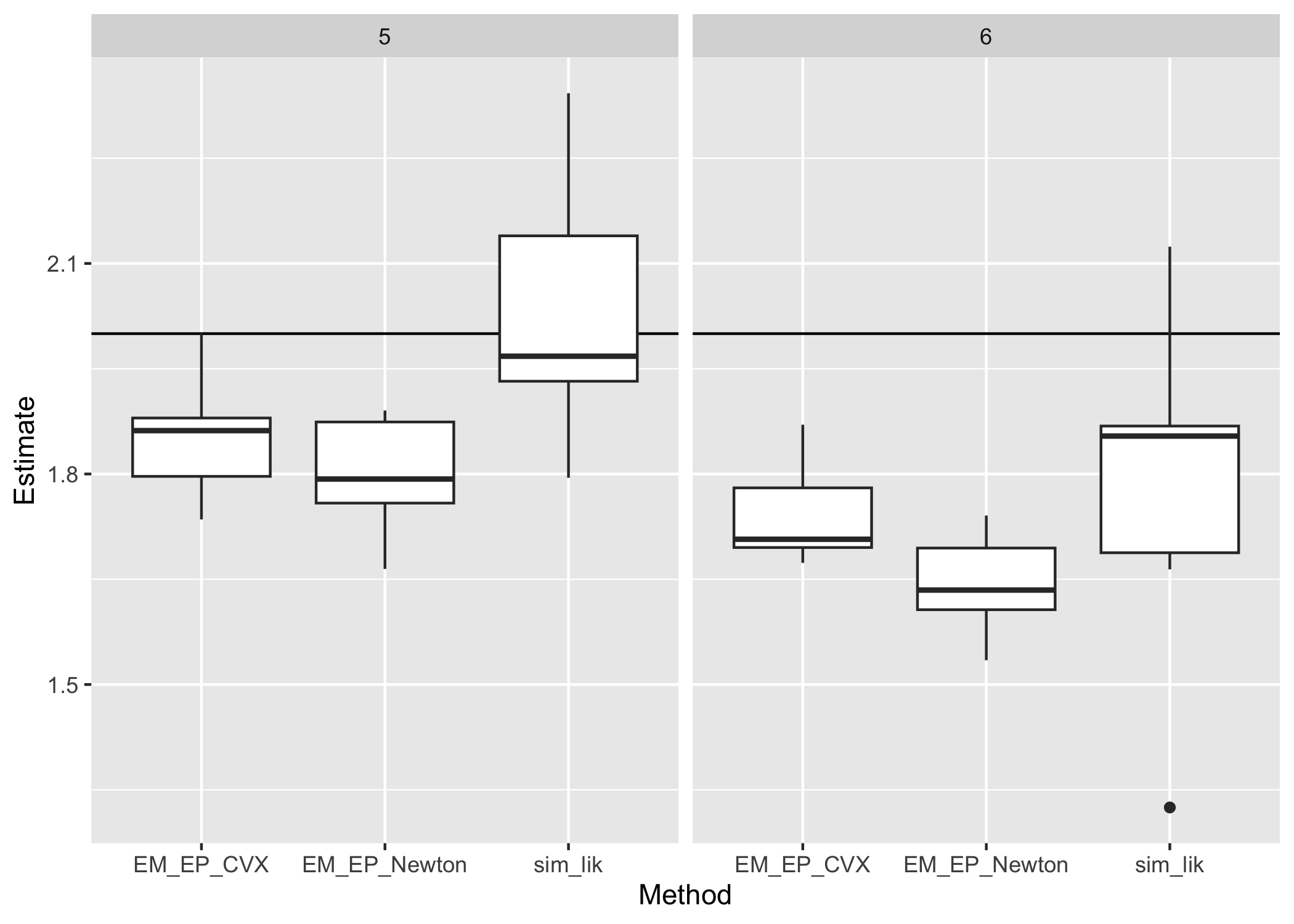}    
            \caption{$\beta$ Error}
        \end{subfigure}
        \\
        \begin{subfigure}[b]{0.32\textwidth}
            \centering
            \includegraphics[width=\textwidth]{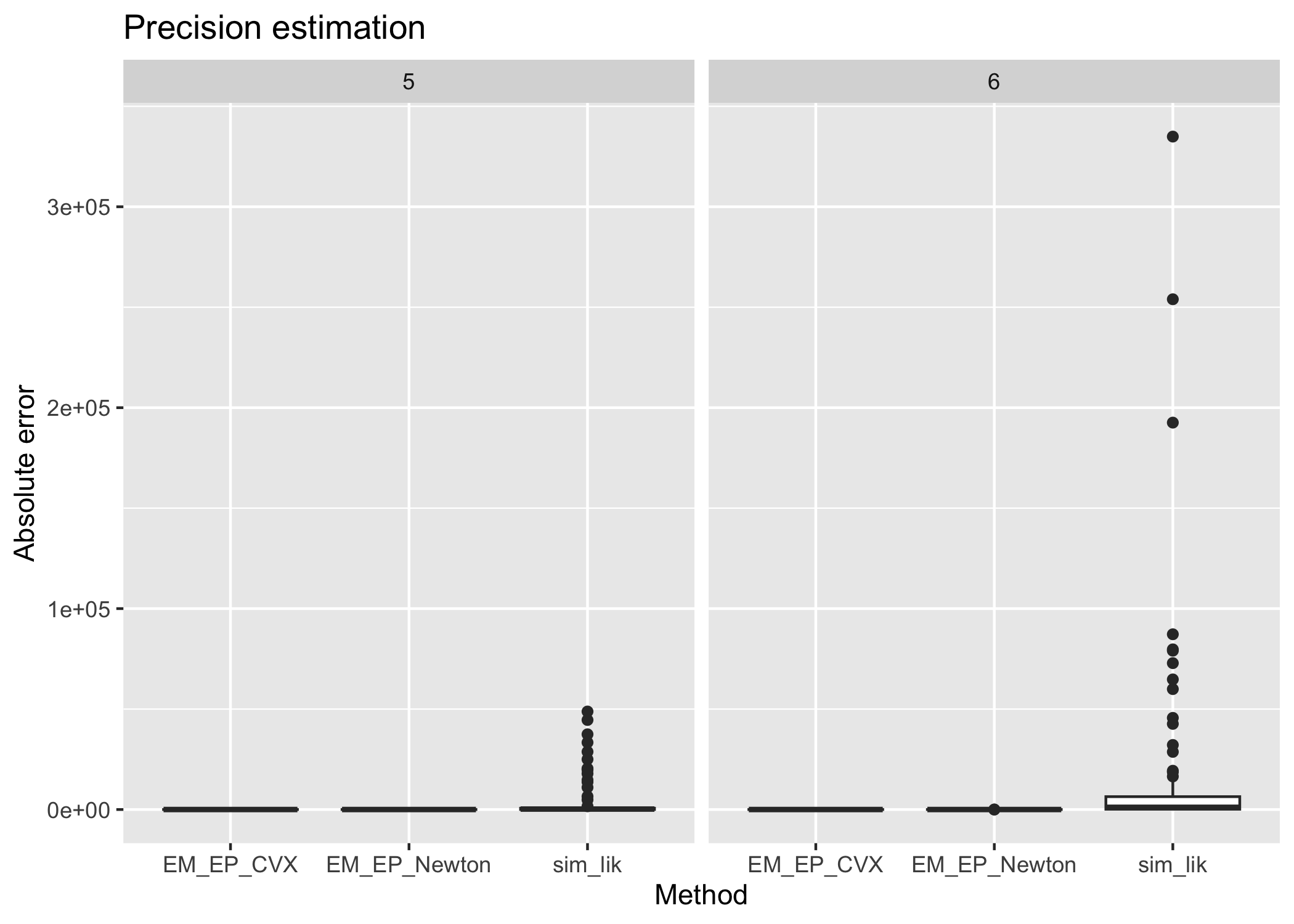}    
            \caption{$\Sigma^{-1}$ Error}
        \end{subfigure}
         \hspace*{.10\textwidth}
        \begin{subfigure}[b]{0.32\textwidth}
            \centering
            \includegraphics[width=\textwidth]{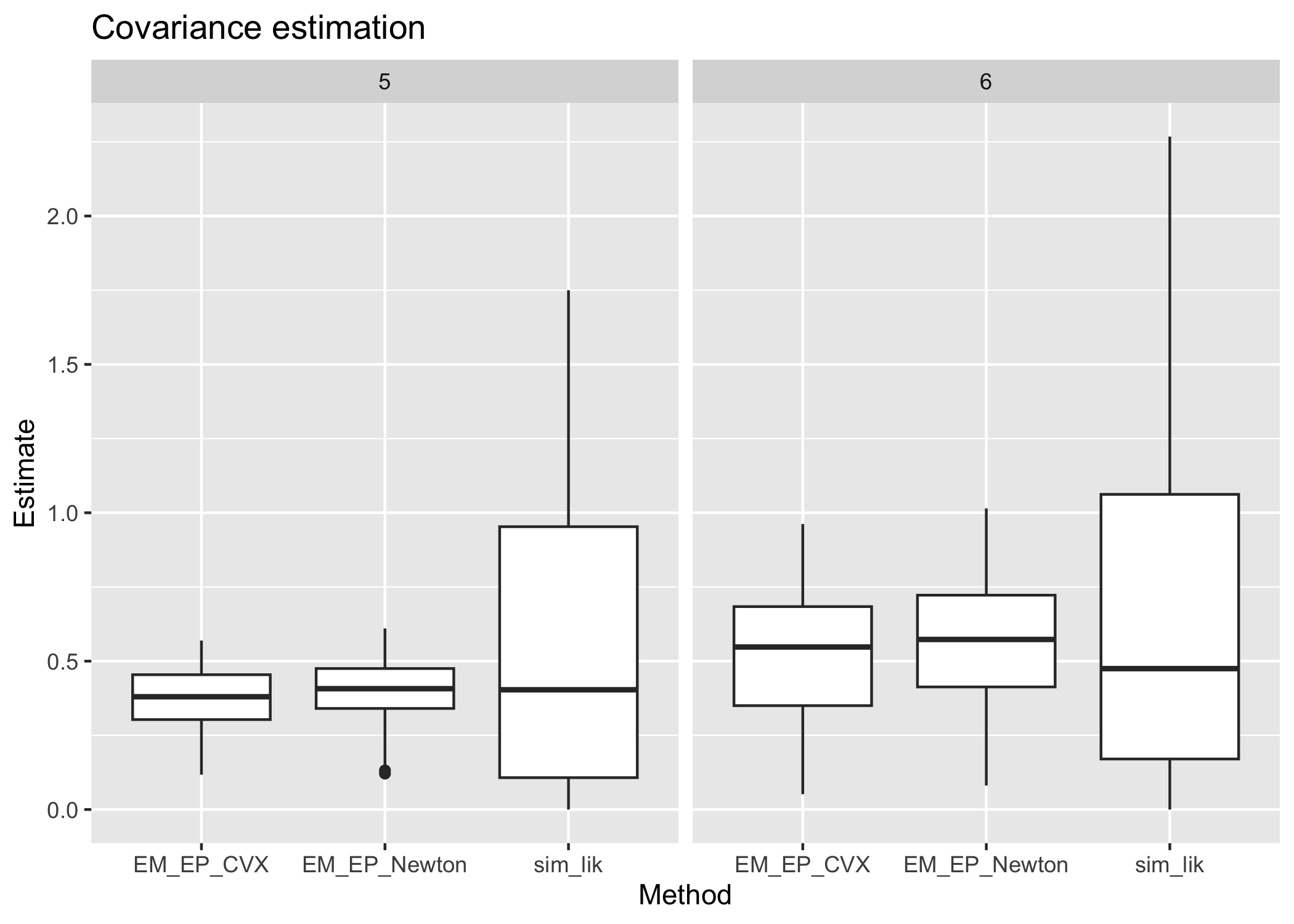}    
            \caption{$\Sigma$ Error}
        \end{subfigure}
        \caption{Comparison of runtime and estimation error of proposed EM algorithm with MC-EM algorithms using Gibbs sampler and Hamiltonian MC sampler. The covariance matrix has a banded structure.}\label{fig:comparison_banded}
    \end{figure}


        \begin{figure}[h!]
        \centering
        \begin{subfigure}[b]{0.32\textwidth}
            \centering
            \includegraphics[width=\textwidth]{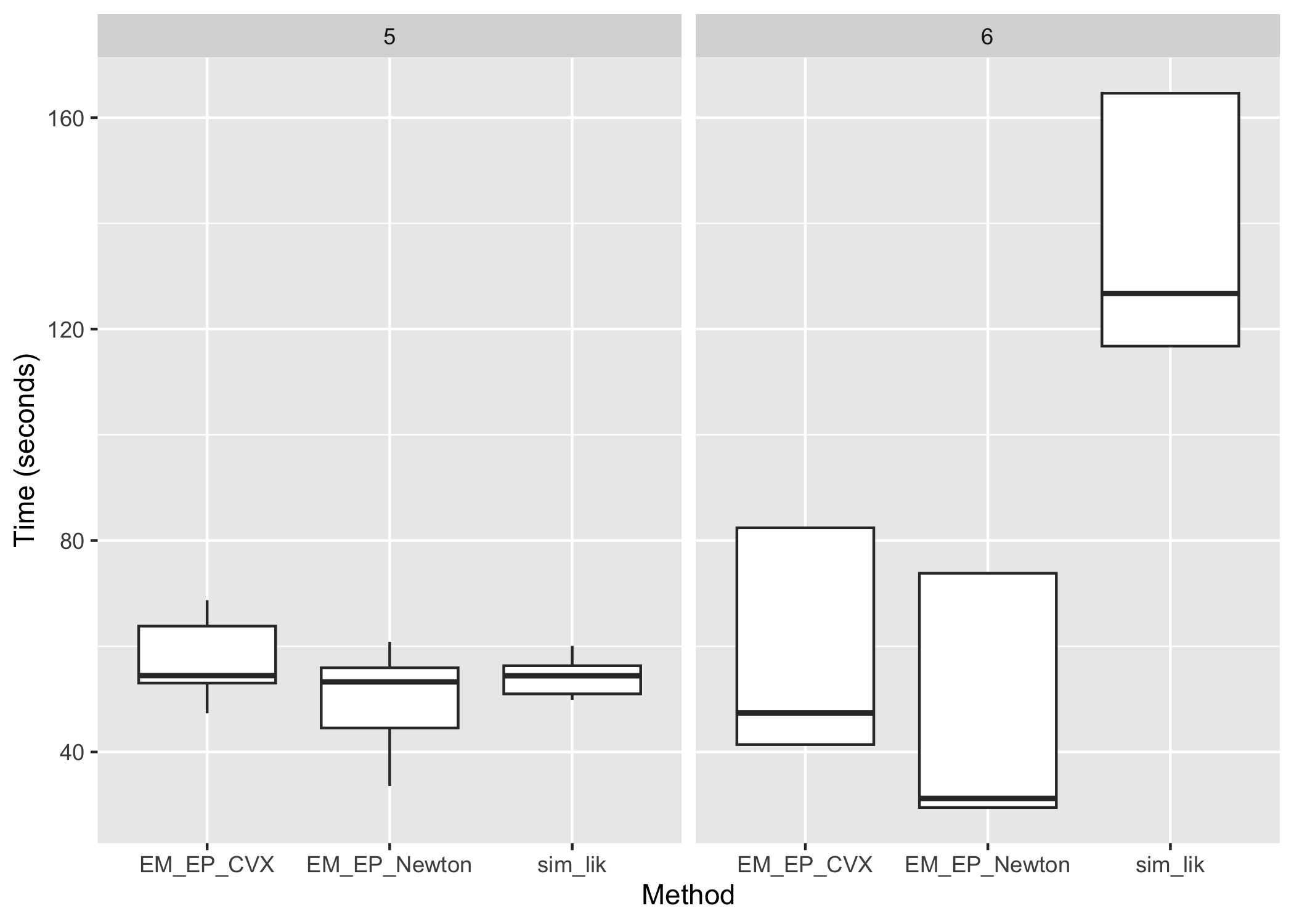}
            \caption{Runtime}
        \end{subfigure}
         \hspace*{.10\textwidth}
        \begin{subfigure}[b]{0.32\textwidth}
            \centering
            \includegraphics[width=\textwidth]{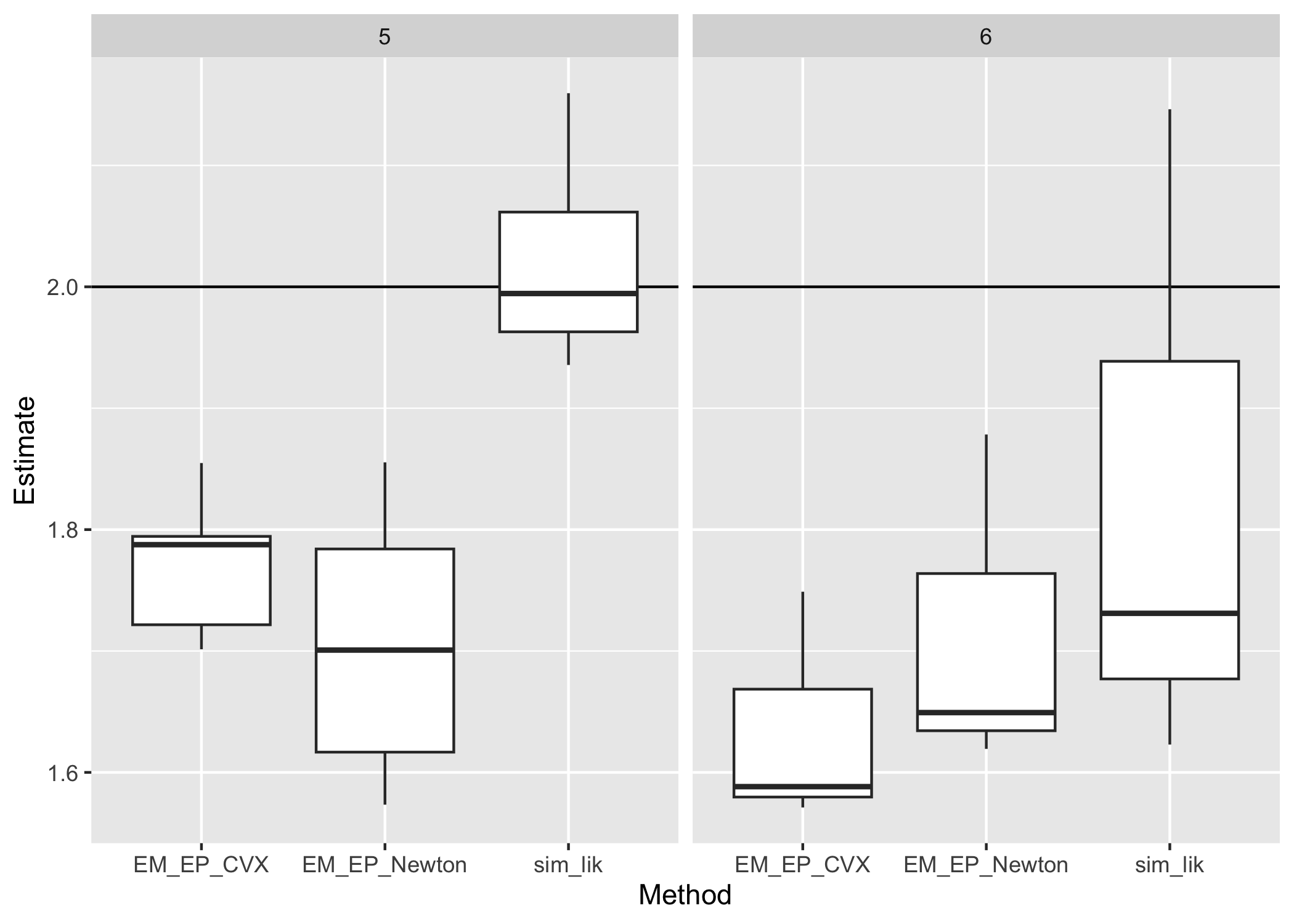}    
            \caption{$\beta$ Error}
        \end{subfigure}
        \\
        \begin{subfigure}[b]{0.32\textwidth}
            \centering
            \includegraphics[width=\textwidth]{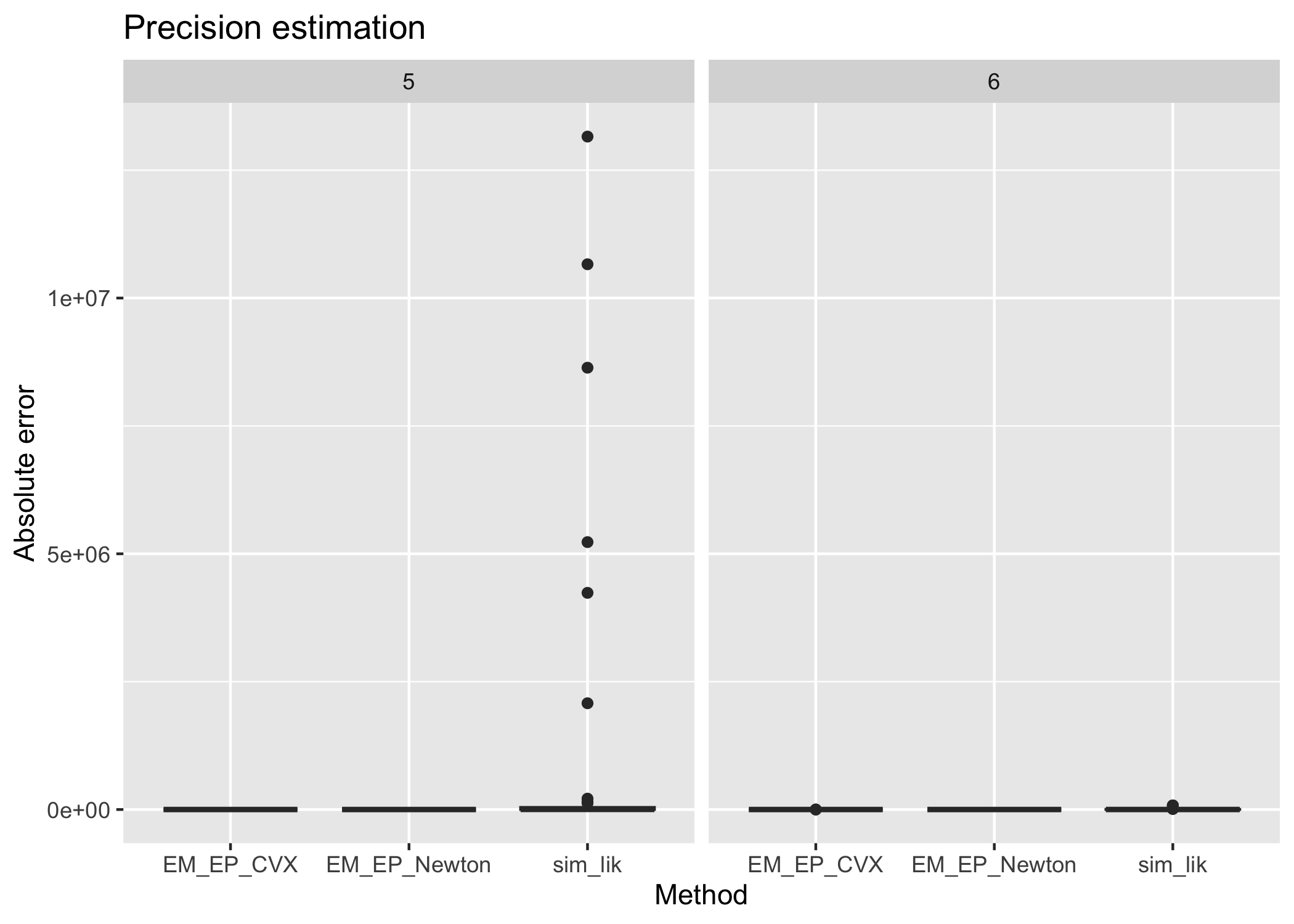}    
            \caption{$\Sigma^{-1}$ Error}
        \end{subfigure}
         \hspace*{.10\textwidth}
        \begin{subfigure}[b]{0.32\textwidth}
            \centering
            \includegraphics[width=\textwidth]{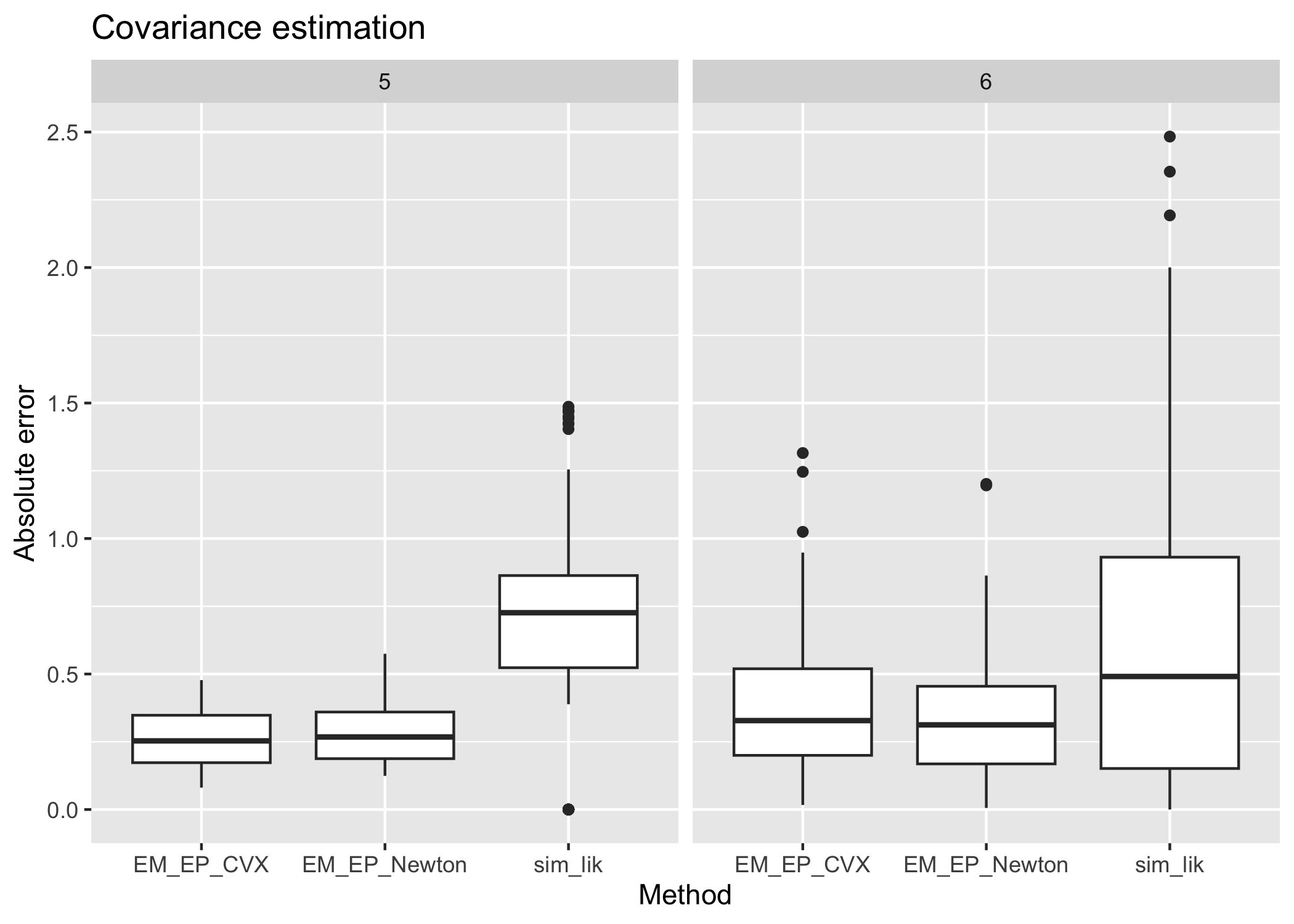}    
            \caption{$\Sigma$ Error}
        \end{subfigure}
        \caption{Comparison of runtime and estimation error of proposed EM algorithm with MC-EM algorithms using Gibbs sampler and Hamiltonian MC sampler. The covariance matrix is randomly generated.}\label{fig:comparison_random}
    \end{figure}
    
\end{appendices}

\end{document}